\documentclass[aps,prb,twocolumn,a4paper,10pt]{revtex4-2}
\usepackage{graphicx}
\usepackage{amsmath}
\usepackage{amssymb}
\usepackage{mathrsfs}
\usepackage{amsfonts}
\usepackage{bm}
\usepackage{mathtools}

\providecommand{\abs}[1]{\left|#1\right|}

\providecommand{\ave}[1]{\langle#1\rangle}
\providecommand{\expe}[1]{\left\langle#1\right\rangle}
\providecommand{\ket}[1]{|#1\rangle}
\providecommand{\bra}[1]{\langle#1|}
\providecommand{\brak}[2]{\langle#1|#2\rangle} 
\providecommand{\proj}[2]{|#1\rangle \! \langle#2|} 
\providecommand{\mean}[3]{\langle#1|#2|#3\rangle} 
\providecommand{\prob}[1]{\operatorname{prob}\left\{#1\right\}} 

\newcommand{\di}{\textrm{d}}
\newcommand{\Hs}{\textrm{H}}
\newcommand{\E}{\textrm{E}}
\newcommand{\bX}{\bm{X}}
\newcommand{\bx}{\bm{x}}
\newcommand{\bmu}{\bm{\mu}}
\newcommand{\hA}{\hat{A}}
\newcommand{\hB}{\hat{B}}
\newcommand{\hC}{\hat{C}}
\newcommand{\hE}{\hat{E}}
\newcommand{\hX}{\hat{X}}
\newcommand{\hY}{\hat{Y}}

\newcommand{\hI}{\hat{I}}
\newcommand{\hU}{\hat{U}}
\newcommand{\hV}{\hat{V}}
\newcommand{\hW}{\hat{W}}

\newcommand{\hO}{\hat{O}}
\newcommand{\hQ}{\hat{Q}}
\newcommand{\hP}{\hat{P}}

\newcommand{\hH}{\hat{H}}
\newcommand{\ha}{\hat{a}}
\newcommand{\had}{\hat{a}^\dagger}
\newcommand{\hadb}{\ha^{\dagger 2}}

\newcommand{\deff}{{\, \vcentcolon = \,}}

\begin{document}


\title{Spectral theorem for dummies: A pedagogical discussion on quantum probability and random variable theory}

\author{Andrea Aiello}
\email{andrea.aiello@mpl.mpg.de} 
\affiliation{Max Planck Institute for the Science of Light, Staudtstrasse 2, 91058 Erlangen, Germany}

\date{\today}

\begin{abstract}
John von Neumann's spectral theorem for self-adjoint operators is a cornerstone of quantum mechanics. Among other things, it also provides a connection between expectation values of self-adjoint  operators and expected values of real-valued random variables.
This paper presents a plain-spoken formulation of this theorem in terms of Dirac's bra and ket notation, which is based on physical intuition and provides techniques that are important for performing actual calculations. The  goal is to engage  students in a constructive discussion about similarities and differences in the use of random variables in classical and quantum mechanics.
Special emphasis is given on  operators that are simple functions of noncommuting self-adjoint operators.
The presentation is self-contained and includes detailed calculations for the most relevant results.
\end{abstract}

\maketitle 

\section{Introduction}\label{Intro}

In the teaching of quantum mechanics it is customary to point out the differences it has with respect to classical mechanics. In contrast, the similarities between the two mechanics are rarely highlighted if at all neglected. Nevertheless, such similarities do exist \cite{Dragoman}, and there are no serious pedagogical reasons for ignoring them. One very important similarity between classical and quantum mechanics concerns the use of random variables.
In  classical mechanics  random variables typically characterize physical observables that have given but \emph{unknown} values. Conversely, in  quantum mechanics random variables describe observables whose values are \emph{uncertain}. In spite of this profound conceptual difference, the actual use of random variables is quite similar in both cases.

The purpose of this work is providing a constructive discussion on the \emph{practical} use of random variables in classical and quantum mechanics, highlighting both similarities and differences. This presentation should help students clarify some seemingly murky and bizarre aspects of quantum mechanics. For this purpose we will  use  von Neumann's spectral theorem  \cite{vonNeumann2018}.
A rigorous mathematical exposition of this theorem is normally not accessible to undergraduate students, because it requires the knowledge of advanced techniques from functional analysis \cite{Riesz_Nagy}. Therefore, here we take a pragmatic approach and present the spectral theorem using  the well-known but non-rigorous Dirac’s methods  \cite{Dirac}. The spirit and level of our presentation strives to conform to that of Sakurai's excellent book on quantum mechanics \cite{Sakurai}.
The emphasis in this paper is on concepts and computations, rather than mathematical rigour  \cite{sep-qt-nvd}, and most calculations are presented in step-by-step detail. As a rule, we leave out the explicit computation of integrals and, instead of providing references to standard tables of integrals, we suggest the reader to use Wolfram Mathematica to perform the calculations \cite{Mathematica}.

The structure of this paper is as follows.
We begin in Sec. \ref{prelude} with a propaedeutical exposition of the spectral theorem for Hermitian matrices of dimension $2 \times 2$.
Then, we continue in Sec. \ref{fundamentals} by presenting a few fundamental concepts and results about random variable theory, which will be used later. In Sec. \ref{ClaQua} we rewrite some key results from random variable theory in a form suitable for showing the connection between classical and quantum probability distributions. Section \ref{Two_examples} gives two examples  in which the previously stated results are applied to the sum and the product of a pair of random variables that follow a normal distribution. So far the presentation has been devoted to the mathematical theory of random variables only. In Sec. \ref{SpecQuant} physics enters presenting the celebrated von Neumann's spectral theorem for self-adjoint operators. Sections \ref{continuous} and \ref{discrete} illustrate the spectral theorem for operators with continuous and discrete spectra, respectively. In Sec. \ref{first} we use the spectral theorem to start a  discussion about quantum and classical probability distributions. The key differences between compatible and incompatible observables in quantum mechanics are also discussed. Section \ref{SumProduct} considers the case of the sum and the product of two noncommuting operators, and introduces the well-known and useful concept of quasi-probability distributions in quantum mechanics. In Sec. \ref{Examples} we consider again the problem studied in Sec. \ref{Two_examples}, but from a quantum-mechanical perspective. This provides for a concrete application of the formulas given in the previous section. Finally, in Sec. \ref{conclusions} we make a few concluding remarks.




\section{Prelude: Spectral decomposition of $2 \times 2$ matrices}\label{prelude}

In a junior-level  algebra course students learn that a square matrix can be written in many different ways. For example, given the matrix
\begin{align}\label{int10}
A = \begin{pmatrix}
      a & b \\
      c & d \\
    \end{pmatrix},
\end{align}
we can write it in the so-called \emph{standard basis} as
\begin{align}\label{int20}
A = a \begin{pmatrix}
      1 & 0 \\
      0 & 0 \\
    \end{pmatrix}
+ b \begin{pmatrix}
      0 & 1 \\
      0 & 0 \\
    \end{pmatrix}
+ c \begin{pmatrix}
      0 & 0 \\
      1 & 0 \\
    \end{pmatrix}
+ d \begin{pmatrix}
      0 & 0 \\
      0 & 1 \\
    \end{pmatrix}.
\end{align}
Alternatively, we could write $A$ in the \emph{Pauli matrices basis} \cite{Sakurai} as
\begin{align}\label{int30}
A = & \; \frac{a + b}{2} \begin{pmatrix}
      1 & 0 \\
      0 & 1 \\
    \end{pmatrix}
+ \frac{b + c}{2} \begin{pmatrix}
      0 & 1 \\
      1 & 0 \\
    \end{pmatrix} \nonumber \\[4pt]
& + i \, \frac{b - c}{2} \begin{pmatrix}
      0 & -i \\
      i & 0 \\
    \end{pmatrix}
+ \frac{a - b}{2} \begin{pmatrix}
      1 & 0 \\
      0 & -1 \\
    \end{pmatrix}.
\end{align}
There is also a third way to write the matrix $A$ when
the latter is Hermitian, that is when $A = A^\dagger$, where $A^\dagger$ denotes the \emph{Hermitian adjoint} of $A$ \cite{horn_johnson_1985}. In this case we can find a set of real eigenvalues $\{a_1, a_2\}$ associated with a set of orthogonal eigenvectors $\{\ket{a_1}, \ket{a_2} \}$, such that
\begin{align}\label{int40}
A \ket{a_i} = a_ i\ket{a_i}, \qquad (i=1,2),
\end{align}
with $\brak{a_i}{a_j} = \delta_{ij}, ~ (i,j=1,2)$. For clarity, let us assume that the eigenvalues are non-degenerate, that is $a_1 \neq a_2$.
Then, it is a simple linear algebra exercise to verify that  $A$ can be written in the \emph{orthogonal projectors basis} $\{ P_1 = \proj{a_1}{a_1}, P_2= \proj{a_2}{a_2} \}$, as
\begin{align}\label{int50}
A =  a_1 P_1 + a_2 P_2,
\end{align}
where $P_i^2 = P_i$, $P_i P_j = P_i \delta_{ij}$ and $P_1 + P_2 = I$, where $I$ is the $2 \times 2$ identity matrix.  For example, if
\begin{align}\label{int60}
A = \begin{pmatrix}
      0 & -i \\
      i & 0 \\
    \end{pmatrix},
\end{align}
then $A^2 = I$, so that its eigenvalues are $a_1 =1$ and $a_2 = -1$.
It is easy to verify via a direct calculation that
\begin{align}\label{int70}
 \begin{pmatrix}
      0 & -i \\
      i & 0 \\
    \end{pmatrix} = (+1)\frac{1}{2} \left( I + A \right)  + (-1) \frac{1}{2} \left( I - A \right),
\end{align}
where $P_1= \left( I + A \right)/2$ and $P_2=\left( I - A \right)/2$ are indeed orthogonal projectors.

Equation \eqref{int50} provides the simplest example of the so-called \emph{spectral decomposition} of a Hermitian matrix.
Such decomposition is not just a mathematical exercise, but has relevant physical consequences. For example, suppose  $A$  represents the projection of the spin of an electron along a fixed direction. Let us call $\sigma$ such observable. Let us also assume that the electron has been prepared in a quantum state represented by the normalized vector $\ket{v}$: $\brak{v}{v}=1$. Then, if we measure $\sigma$ on a collection of many identically prepared electrons, its  measured mean value $\expe{\sigma}_\text{measured}$,  will be given by $\expe{\sigma}_\text{measured} \approx \mean{v}{A}{v}$, where
\begin{align}\label{int75}
\mean{v}{A}{v}  & = \mean{v}{a_1 P_1 + a_2 P_2}{v} \nonumber \\[4pt]
& =  a_1 \mean{v}{ P_1}{v} + a_2 \mean{v}{P_2}{v} \nonumber \\[4pt]
& =  a_1 \abs{\brak{a_1}{v}}^2 + a_2 \abs{\brak{a_2}{v}}^2 \nonumber \\[4pt]
&  \deff   a_1 w_1 + a_2 \, w_2.
\end{align}
All the quantities in \eqref{int75} have a clear physical meaning: $a_1$ and $a_2$ are the
values that $\sigma$ may take when measured, and $w_i$ is the relative frequency of occurrence of  $a_i$ when several measurements of $\sigma$ are made. This interpretation makes sense because by definition the $w_i = \abs{\brak{a_i}{v}}^2$ are nonnegative  and add to  $1$:
\begin{align}\label{int77}
w_1 + w_2 = & \; \mean{v}{ P_1}{v} +  \mean{v}{P_2}{v} \nonumber \\[4pt]
= & \;  \mean{v}{ P_1 +  P_2}{v} \nonumber \\[4pt]
= & \; \brak{v}{v} = 1.
\end{align}
Therefore, we may infer that $w_i$ gives the probability that a measurement of $\sigma$ yields the value $a_i$. Thus, the importance of the spectral decomposition \eqref{int50}
 over the other ways of writing the matrix $A$ lies in the fact that it provides for both the values
of the observable represented by $A$, and the probability that each of such values occur  when the observable is measured.

In quantum mechanics we often deal with spaces of infinite dimensions, where the concept of  matrix as representation of linear transformations  is replaced  by the more general concept of linear operator. Moreover, while every matrix on a finite dimensional space has a discrete spectrum and is bounded, a linear operator in an infinite-dimensional space may possess a continuous spectrum and also be unbounded. Therefore, rather than dealing with linear operators with discrete and continuous spectra separately,  it is more convenient to rewrite the spectral decomposition  \eqref{int50} in a slightly different form that includes both cases. For this purpose we introduce the concept of  \emph{spectral family} generated by $A$, denoted $E_A(\lambda)$, and defined by
\begin{align}\label{int80}
E_A(\lambda) \deff  P_1 \, \Hs(\lambda - a_1) + P_2 \, \Hs(\lambda - a_2),
\end{align}
where  $\lambda$ is a real parameter and $\Hs(x)$ is the Heaviside step function \cite{lighthill_1958,Hfunction}. From the definition \eqref{int80} it follows that
\begin{align}\label{int90}
E_A(\lambda) = \left\{
                 \begin{array}{ll}
                   O , & \lambda \leq \min(a_1,a_2) , \\[4pt]
                   P_1 , & \min(a_1,a_2) < \lambda \leq \max(a_1,a_2) , \\[4pt]
                   P_1 + P_2 = I, & \lambda > \max(a_1,a_2) .
                 \end{array}
               \right.
\end{align}
where $O $ denotes a $2 \times 2$ matrix that has all its elements equal to zero.
Using $\di \Hs(x)/\di x = \delta(x)$, where $\delta(x)$ is the Dirac delta distribution \cite{DeltaFunction}, we can
rewrite \eqref{int50} as
\begin{align}\label{int100}
A  = & \; a_1 P_1 + a_2 P_2 \nonumber \\[4pt]
= & \; \int_{-\infty}^\infty \lambda \bigl[ P_1 \, \delta(\lambda - a_1) + P_2 \, \delta(\lambda - a_2) \bigr]\, \di \lambda \nonumber \\[4pt]
= & \;  \int_{-\infty}^\infty \lambda \, \frac{\di E_A(\lambda)}{\di \lambda} \, \di \lambda  \nonumber \\[4pt]
= & \; \int_O^{I} \lambda \, \di E_A(\lambda),
\end{align}
where the last integral must be understood in the sense of a \emph{Stieltjes integral} \cite{Stieltjes}, and \eqref{int90} has been used.

It was a remarkable achievement of  von Neumann to have shown that in infinite-dimensional space there exist unbounded self-adjoint operators for which the decomposition \eqref{int100} is still valid \cite{vonNeumann2018}. This result is known in quantum mechanics as ``\emph{the spectral theorem}''. In the following, after a short review of random variable theory, we will present this theorem in a mathematically non-rigorous way and, more importantly, explore its physical consequences.

\section{A few facts about  random variable theory}\label{fundamentals}

In this section, we briefly review some key concepts in probability and random variables theory that we will use later. The principal references  we used to prepare this section are the  books by Feller \cite{FellerII}, Papoulis\&Pillai \cite{papoulis2002}, and  Taboga \cite{taboga2017}, and the articles by  Gillespie  \cite{Gillespie1983}, Ramshaw \cite{Ramshaw1985}, and Paveri-Fontana \cite{Paveri-Fontana1991}.

\subsection{One random variable}\label{oneRV}

In everyday life, we often observe things happening in an apparently unpredictable way, as the variations of the stock market prices, or the result of a sport game, or  also the results of a measurement in a quantum physics experiment. Henceforth, by the word ``experiment'' we will denote the \emph{observation} and \emph{recording} of some thing happening. The way the observed thing turns out is called an ``outcome'' or a ``result'' of the experiment.
The set of all possible  outcomes  of an experiment, here recorded and labeled as $\omega,~ \omega',~ \omega'',\dots,$ is called the \emph{sample space} of the experiment, and it is usually denoted by $\Omega = \left\{ \omega,~ \omega',~ \omega'',\dots \right\}$. A \emph{random variable} $X$ is a  function from the sample space $\Omega$ to the set of  real numbers $\mathbb{R}$, that assigns a real number $x$ to every  outcome $\omega$ of an experiment, that is
\begin{align*}
X  :  \Omega \rightarrow \mathbb{R} \, .
\end{align*}
Here and hereafter we will use capital and lowercase letters, as $X$ and $x$, to distinguish a random variable $X$, from the value $x$ that it may take, so that $x = X(\omega), ~ x' = X(\omega')$, et cetera.

When $X$   takes a countable number  of values, it  is called a  discrete random variable. Conversely, if the image of $X$ is uncountably infinite, like an interval of $\mathbb{R}$, then $X$ is called a continuous random variable. In both cases, the probability that  $X$ takes a value less or equal to $x$, is called the cumulative distribution function (CDF) $F_X(x)$ of $X$:
\begin{align}\label{a20}
F_X(x) \deff  \prob{X \leq x} = \int_{-\infty}^x f_X(x') \, \di x',
\end{align}
where ``prob'' stands for probability, and
\begin{align}\label{a30}
f_X(x) \, \di x = \di F_X(x),
\end{align}
is  the probability density function (PDF)  of $X$.
By definition, $f_X(x)$ is normalized according to
\begin{align}\label{a20bis}
F_X(\infty) \deff  \prob{X < \infty} = \int_{-\infty}^\infty f_X(x') \, \di x' = 1,
\end{align}
because every time  we run the experiment one possible outcome $\omega$ will certainly occur,  so that the probability that $X$ will take any value $-\infty < x = X(\omega) < \infty$, will be equal to one.

While $F_X(x)$ is always an ordinary function, $f_X(x)$ is not when $X$ is a discrete random variable. Suppose that $X$    takes the denumerable values $x_m$ with probability $p_m \geq 0$, where $m = 1,2, \ldots, M$, with $M \in \mathbb{N}$ either finite or infinite (here and hereafter the symbol $\mathbb{N}$ denotes the natural numbers $\{1,2,3, \ldots \}$). Then,   $f_X(x)$ is a \emph{generalized function} defined by
\begin{align}\label{a40}
f_X(x) = \sum_{m=1}^{M} p_m  \delta \left( x - x_m \right),
\end{align}
where $\delta(x)$ is the Dirac delta distribution \cite{DeltaFunction}, and
\begin{align}\label{a50}
\sum_{m=1}^{M} p_m = 1.
\end{align}
The probability $p_m$ is often written in terms of the so-called \emph{probability mass function} $p_X(x)$, defined by
\begin{align}\label{a52}
p_X(x) = \prob{X = x},
\end{align}
so that $p_m = p_X(x_m)$, and
\begin{align}\label{a54}
f_X(x) = \sum_{m=1}^{M} p_X(x) \, \delta ( x - x_m ),
\end{align}
Next, using \eqref{a20} we readily find
\begin{align}\label{a60}
F_X(x) = & \; \sum_{m=1}^{M} \int_{-\infty}^x  p_X(x') \, \delta ( x' - x_m ) \, \di x' \nonumber \\[4pt]
= & \; \sum_{m=1}^{M} p_m \, \Hs ( x - x_m ) ,
\end{align}
where $\Hs(x)$ is the Heaviside step function \cite{lighthill_1958,Hfunction}. This shows that although $f_X(x)$ is a generalized function, $F_X(x)$ is an ordinary function, even if noncontinuous.

The \emph{expected value} of $X$, denoted as either $\E \left[ X \right]$ or $\expe{X}$, is given by
\begin{align}\label{a70}
\expe{X} = \int_{\mathbb{R}}  x \, f_X(x) \, \di x.
\end{align}
Let $g(x)$ be a real-valued function of the real variable $x$: $g  :  \mathbb{R} \rightarrow \mathbb{R}$.  The expected value of the new random variable $G = g(X)$, is given by
\begin{align}\label{a80}
\expe{g(X)} =   \int_{\mathbb{R}}  g(x) \, f_X(x) \, \di x.
\end{align}
This  and \eqref{a20bis} imply
\begin{align}\label{a80new}
\expe{\alpha + \beta \, g(X)} =  \alpha + \beta \expe{g(X)}, \qquad (\alpha ,\beta \in \mathbb{C}).
\end{align}

If $\zeta(x)$ is a complex-valued function of $x \in \mathbb{R}$, that is $\zeta  :  \mathbb{R} \rightarrow \mathbb{C}$, then \eqref{a80} is still valid because we can always write $\zeta(x) = u(x) + i \, v(x)$, where $u(x)$ and $v(x)$ are real-valued functions, so that \eqref{a80} gives
\begin{align}\label{a81}
\expe{\zeta(X)} =   \expe{u(X)}  + i \expe{v(X)} .
\end{align}
A very useful complex-valued function is $\zeta(x) = \exp( i s x)= \cos(  s x) + i \sin (s x) $, with $s,x \in \mathbb{R}$. In this case
\begin{align}\label{a82}
\expe{\exp( i s X)} = \int_{\mathbb{R}}  e^{i s x} \, f_X(x) \, \di x \deff  \varphi_X(s) ,
\end{align}
where $\varphi_X(s)$ is the so-called \emph{characteristic function} of the random variable $X$. By definition, $\varphi_X(s)$ is the Fourier transform of $f_X(x)$ when the latter exists. When $\varphi_X(s)$ is known,  then  $f_X(x)$ is univocally determined via Fourier inversion, and vice versa:
\begin{align}\label{a83}
f_X(x) = \frac{1}{2 \pi} \int_{\mathbb{R}}  e^{-i s x} \, \varphi_X(s) \, \di s.
\end{align}
 Characteristic functions are valuable because they permit to derive important results about moments and sums of random variables without knowing explicitly their PDFs \cite{FellerII}. For example, it follows from \eqref{a80} and \eqref{a82} that the $n^\text{th}$ moment of a random variable $X$ distributed according to $f_X(x)$ is given by
\begin{align}\label{a84}
\frac{1}{i^n}  \left. \frac{\di^n \varphi_X(s)}{\di s^n} \right|_{s=0} = \int_{\mathbb{R}}  x^n  f_X(x) \, \di x.
\end{align}
We leave it to the reader to calculate the characteristic function of a discrete random variable and verify that in this case \eqref{a83} correctly produces \eqref{a40}.

\subsection{Two or more random variables}\label{vectorRV}

All the concepts above generalize directly to the case of a set of two or more random variables, that is the so-called
vector random variables, or random vectors for short. A \emph{random vector} $\bX = (X_1, X_2, \ldots, X_N)$ is a  function from the sample space $\Omega$ to the set of $N-$dimensional real vectors $\mathbb{R}^N$, that assigns a real vector $\bx = (x_1, x_2, \ldots, x_N)$ to every  outcome $\omega$ of an experiment, that is
\begin{align*}
\bX  :  \Omega \rightarrow \mathbb{R}^N \, .
\end{align*}
The \emph{joint cumulative distribution function}  $F_{X_1 \cdots X_N}(\bx)=F_{X_1 \cdots X_N}(x_1, \ldots, x_N)$ of $\bX$, is defined as
\begin{multline}\label{a100}
F_{X_1 \cdots X_N}(\bx) \\[4pt] =  \prob{X_1 \leq x_1, ~ X_2 \leq x_2, \ldots, X_N \leq x_N} \\[4pt] =
 \int_{-\infty}^{x_N} \cdots \int_{-\infty}^{x_1} f_{X_1 \cdots X_N}(\bx') \, \di x_1' \cdots \di x_N',
\end{multline}
where
\begin{align}\label{a110}
f_{X_1 \cdots X_N}(\bx) = & \;  f_{X_1 \cdots X_N}(x_1, \ldots, x_N)\nonumber \\[4pt]
 = & \;   \frac{\partial^N F_{X_1 \cdots X_N}(x_1, \ldots, x_N)}{\partial x_1 \cdots \partial x_N} \, ,
\end{align}
is  the \emph{joint probability density function}   of $\bX$.

Let $g(\bx)=g(x_1, \ldots, x_N)$ be an ordinary function of the real $N$-dimensional vector $\bx$: $g  :  \mathbb{R}^N \rightarrow \mathbb{R}$.  The expected value of $ g(\bX)$ is
\begin{align}\label{a120}
\expe{g(\bX)} =   \int_{\mathbb{R}^N}  g(\bx) \, f_{X_1 \cdots X_N}(\bx) \, \di^N x \, ,
\end{align}
where $\di^N x = \di x_1 \cdots \di x_N$. If $h(x_1, \ldots, x_K)$ is a real function from $\mathbb{R}^K$ to $\mathbb{R}$,  with $K < N$, then \eqref{a120} implies
\begin{align}\label{a130}
\langle h(X_1,   \ldots, & \, X_K \rangle \nonumber
\\[4pt]
  =  & \;  \int_{\mathbb{R}^N}  h(x_1, \ldots, x_K) \, f_{X_1 \cdots X_N}(\bx) \, \di^N x
\nonumber
\\[4pt]  =  & \; \int_{\mathbb{R}^K}  h(x_1, \ldots, x_K) \Biggl\{ \int_{\mathbb{R}^{N-K}}  f_{X_1 \cdots X_N}(\bx)
\nonumber \\[4pt]  & \times  \, \di x_{K+1} \cdots \di x_{N} \Biggr\}  \, \di x_1 \ldots \di x_K .
\end{align}

This equation may be rewritten as
\begin{multline}\label{a140}
\expe{h(X_1, \ldots, X_K)} \\[4pt]
=  \int_{\mathbb{R}^K}  h(x_1, \ldots, x_K) \,f_{X_1 \cdots X_K}(x_1, \ldots, x_K) \, \di^K x,
\end{multline}
where $\di^K x = \di x_1 \cdots \di x_K$, and we have introduced the \emph{marginal} probability density function $f_{X_1 \cdots X_K}(x_1, \ldots, x_K)$, defined as

\begin{multline}\label{a150}
f_{X_1 \cdots X_K}(x_1, \ldots, x_K) \\[4pt]
\deff   \int_{\mathbb{R}^{N-K}}  f_{X_1 \cdots X_N}(x_1,  \ldots, x_N) \, \di x_{K+1} \cdots \di x_N.
\end{multline}

\section{Some fundamental results}\label{ClaQua}

The following results are key  to establish a connection between classical and quantum probability distributions.
From the trivial identity
\begin{align}\label{a80b}
g(x')  = \int_{\mathbb{R}}  g(x) \, \delta \left( x - x' \right) \, \di x,
\end{align}
it follows the less trivial relation
\begin{align}\label{a80c}
\expe{g(X)} = & \;    \int_{\mathbb{R}}  g(x') f_X (x') \, \di x' \nonumber \\[4pt]
= & \;    \int_{\mathbb{R}}  \left[ \int_{\mathbb{R}} g(x) \delta (x - x') \di x \right] f_X (x') \, \di x' \nonumber \\[4pt]
= & \;    \int_{\mathbb{R}} g(x)  \left[ \int_{\mathbb{R}}  \delta (x - x') f_X (x') \di x' \right]   \di x \nonumber \\[4pt]
= & \;    \int_{\mathbb{R}} g(x)  \expe{\delta (x - X)}   \di x ,
\end{align}
where \eqref{a80} has been used and we passed from the second to the third line by interchanging $x$ and $x'$ integrations.
By equating the right side of \eqref{a80} with the right-hand side of \eqref{a80c}, we obtain
\begin{align}\label{a80d}
\int_{\mathbb{R}}  g(x) \, f_X(x) \, \di x = \int_{\mathbb{R}}  g(x) \expe{\delta \left( x - X  \right)}  \di x .
\end{align}
Since this  must hold  for any $g(x)$, we infer the important relation \cite{Ramshaw1985},
\begin{align}\label{a80e}
 f_X(x) = \expe{\delta \left( x - X  \right)}  \, .
\end{align}
Here and hereafter, when we write an expression like this, we always intend that the expected value should  be taken with respect to the PDF of the random variable shown in the equation (i.e., $X$ in this case).

It is not difficult to show that \eqref{a80e} is fully consistent with \eqref{a20} and \eqref{a30}. Since by definition
\begin{align}\label{a80f}
 \int_{-\infty}^x f_X(x') \, \di x' =  & \;  \int_{\mathbb{R}} \Hs (x - x') f_X(x') \, \di x' \nonumber \\[4pt]
= & \; \expe{\Hs (x - X)},
\end{align}
then from \eqref{a20} it follows that \cite{Paveri-Fontana1991},
\begin{align}\label{a80g}
F_X(x) = \expe{\Hs (x - X)}.
\end{align}
Taking the derivative of both sides of this equation with respect to $x$, and using \eqref{a30}, we obtain again \eqref{a80e}.

The validity of \eqref{a80e} can be also verified using the integral representation of the delta distribution
\begin{align}\label{a80h}
\delta \left( x - y \right)  = \frac{1}{2 \pi} \int_{\mathbb{R}}  e^{-i s \left( x - y \right)} \,  \di s ,
\end{align}
and the characteristic function of $X$. Substituting the left-hand side of \eqref{a82} into \eqref{a83}, and using \eqref{a80new}, we obtain
\begin{align}\label{a80i}
f_X(x) = & \; \frac{1}{2 \pi} \int_{\mathbb{R}}  e^{-i s x} \expe{e^{i s X}}  \di s \nonumber \\[4pt]
= & \; \expe{\frac{1}{2 \pi} \int_{\mathbb{R}}  e^{-i s (x-X)}  \di s} \nonumber \\[4pt]
= & \; \expe{\delta \left( x - X \right)} ,
\end{align}
which coincides with \eqref{a80e}.

The key results \eqref{a80e} and \eqref{a80g} can be directly generalized to the case of two or more random variables. It is a simple exercise to show that it is always possible to write
\begin{align}
f_{X_1 \cdots X_N}(\bx) = & \; \expe{\delta \left( \bx - \bX  \right)}   , \label{a160} \\[4pt]
F_{X_1 \cdots X_N}(\bx) = & \;  \expe{\Hs (\bx - \bX)}   , \label{a170}
\end{align}
where
\begin{align}
\delta \left( \bx - \bX  \right) = & \; \prod_{n=1}^N \delta \left( x_n - X_n  \right)   , \label{a180} \\[4pt]
\Hs (\bx - \bX) = & \;  \prod_{n=1}^N \Hs \left( x_n - X_n  \right)  . \label{a190}
\end{align}

A straightforward byproduct of \eqref{a160} is the so-called \emph{random variable transformation} (RVT) \emph{theorem}. Any real-valued function of random variables, say $g(\bX)=g(X_1, \ldots, X_N)$, is itself a random variable, say $Y = g(X_1, \ldots, X_N)$. The RVT theorem states that the PDF $f_Y(y)$ of $Y$ is
\begin{align}\label{a192}
f_Y(y) = & \; \expe{\delta  \bigl(y - g\left(X_1, \ldots, X_N  \right) \bigr)} \nonumber \\[4pt]
= & \; \int_{\mathbb{R}^N}  \delta \bigl(y - g\left( \bx  \right) \bigr) \, f_{X_1 \cdots X_N}(\bx) \, \di^N x,
\end{align}
where $\bx = (x_1,  \ldots, x_N)$. The straightforward proof of this theorem, can be found in \cite{Gillespie1983,Ramshaw1985}.

The integral in \eqref{a192} can be calculated in two different ways. In the first way we choose one of the $N$ variables $\{ x_1, \ldots, x_N\}$, say $x_n$, and consider the quantity $y - g\left( \bx  \right)$ as a function of $x_n$: $h(x_n) \deff  y - g\left( x_1, \ldots, x_n , \ldots,x_N  \right)$. We then apply the formula for the delta distribution of a  function of $x_n$, which is given by \cite{DeltaFunction}:
\begin{align}\label{delta10}
\delta \bigl( h(x_n) \bigr) =  \sum_{r =1}^R \frac{ \delta \bigl( x_n - x_n^{(r)} \bigr) }{\abs{\left.\frac{\di h (x_n)}{\di x_n}\right|_{x_n = x_n^{(r)}}}} ,
\end{align}
where $\{ x_n^{(1)},x_n^{(2)}, \ldots, x_n^{(R)}\}$ are the supposedly $R$ roots of the equation $h(x_n) = 0$. These roots, by definition, depend on both the variable $y$ and the remaining $N-1$  variables $\{x_1, \ldots, x_{n-1}, x_{n+1}, \ldots, x_N \}$. Substituting the right-hand side of \eqref{delta10} into the right-hand side of \eqref{a192}, we can perform the $x_n$ integration trivially using \eqref{delta10}, and then we must hope that the remaining $N-1$ integrals can be feasible.
Of course, an either judicious or unwise choice of $x_n$ may make it easier or more difficult to calculate the remaining $N-1$ integrals.

The second method of performing the integral in \eqref{a192} is more elegant and is based on the integral representation \eqref{a80h} of the delta distribution and on the definition of characteristic function \eqref{a82}. Using \eqref{a80h}  into \eqref{a192} and interchanging the $x$ and $s$ integrations, we get
\begin{align}\label{delta20}
f_Y(y) = & \; \frac{1}{2 \pi} \int_{\mathbb{R}^{N+1}}  e^{-i s \left(  y- g\left( \bx  \right)  \right)}  \, f_{X_1 \cdots X_N}(\bx) \, \di s \, \di^N x \nonumber \\[4pt]
= & \;  \frac{1}{2 \pi}  \int_{\mathbb{R}} e^{-i s y} \varphi_Y(s)   \, \di s,
\end{align}
where by definition
\begin{align}\label{delta30}
\varphi_Y(s) =   \int_{\mathbb{R}^{N}}  e^{i s  g\left( \bx  \right) }  \, f_{X_1 \cdots X_N}(\bx) \, \di^N x ,
\end{align}
is the characteristic function of $Y$.
In the next section  we will use both methods of integration for illustrative purposes.

\section{Two elementary examples}\label{Two_examples}

The following two applications of the formulas given in the previous sections, are trivial from the point of view of random variable theory. However, as we will see later, they become challenging in the framework of quantum mechanics. Therefore, it is instructive to present them first in detail here. In both examples $\bX = (X_1,X_2)$ is a two-dimensional Gaussian random vector with means $\bmu = (\mu_1, \mu_2)$, and covariance matrix $V$ defined by
\begin{align}\label{e10}
V = \begin{bmatrix}
      \sigma_1^2 & \rho \, \sigma_1 \sigma_2 \\[4pt]
      \rho \, \sigma_1 \sigma_2 & \sigma_2^2 \\
    \end{bmatrix} ,
\end{align}
with $\sigma_1, \sigma_2 >0$, and $-1 \leq \rho \leq 1$.
It is customary to write either $\bX \sim \mathcal{N}_2(\bmu,V)$ or $\bX \sim f_{X_1 X_2}(x_1,x_2)$,  to indicate that the random vector $\bX$ follows a normal bivariate distribution characterized by $\bmu $ and  $V$ \cite{papoulis2002}, where
\begin{align}\label{e20}
f_{X_1 X_2}(x_1,x_2) = \, &  \frac{1}{2 \pi \, \sqrt{\det V}} \nonumber \\[4pt]
\, & \times \exp \left[- \frac{1}{2} (\bx - \bmu)^T V^{-1}  (\bx - \bmu)\right],
\end{align}
with $\bx = (x_1,x_2)$, and
\begin{equation}
\begin{split}\label{e20ter}
 \mu_n \deff  & \;  \expe{X_n} =  \int_{\mathbb{R}^2}  x_n \, f_{X_1 X_2}(x_1,x_2) \, \di x_1 \di x_2 ,  \\[4pt]
V_{nm} \deff  & \;  \expe{(X_n - \mu_n) (X_m - \mu_m)}\\[4pt]
= & \;   \int_{\mathbb{R}^2}  (x_n - \mu_n) (x_m - \mu_m) \, f_{X_1 X_2}(x_1,x_2) \, \di x_1 \di x_2 ,
\end{split}
\end{equation}
where $n,m = 1,2$.

From \eqref{a150} and a straightforward integration, it follows that the marginal distributions $f_{X_1 }(x_1)$ and $f_{X_2}(x_2)$, are given by
\begin{equation}\label{e20bis}
\begin{split}
f_{X_1 }(x_1) = & \; \frac{1}{\sqrt{2 \pi } \, \sigma_1} \exp\left[-\frac{\left(x_1 - \mu_1\right)^2}{2 \, \sigma_1^2}\right],  \\[4pt]
f_{X_2}(x_2) = & \;  \frac{1}{\sqrt{2 \pi } \, \sigma_2}  \exp\left[-\frac{\left(x_2 - \mu_2\right)^2}{2 \, \sigma_2^2}\right],
\end{split}
\end{equation}
so that $X_n \sim \mathcal{N}_1(\mu_n, \sigma_n^2),~ (n=1,2)$.

\subsection{Sum of two random variables}\label{ExampleSum}

Consider  the \emph{sum} of  $X_1$ and $X_2$:   $g(X_1,X_2) = X_1 + X_2$. From \eqref{a120} it follows that
\begin{align}\label{e90}
\expe{X_1+  X_2} = & \;  \int_{\mathbb{R}^2}  \left(x_1 + x_2 \right) \, f_{X_1 X_2}(x_1,x_2) \, \di x_1 \di x_2 \nonumber \\[4pt]
= & \; \mu_1 + \mu_2.
\end{align}
However, we could also consider $X_1 + X_2$ as a single random variable, say $W = X_1 + X_2$. By hypothesis, $W \sim f_W(w)$, so that we could calculate $\expe{W}$ not only from\eqref{e90}, but also as
\begin{align}\label{e40}
\expe{W} = \int_{\mathbb{R}}  w \, f_W(w) \, \di w.
\end{align}
To find $f_W(w)$ we use the RVT theorem  \eqref{a192} with $g(x_1,x_2) = x_1 + x_2$. We  apply the first method of integration choosing $x_2$ as first integration variable and find
\begin{align}\label{e100}
f_{W}(w)
= & \; \expe{\delta  \bigl(w - X_1 - X_2   \bigr)} \nonumber \\[4pt]
= & \; \int_{\mathbb{R}} f_{X_1 X_2} \left( x_1, w - x_1 \right) \, \di x_1 \nonumber \\[4pt]
= & \; \frac{1}{\sqrt{2 \pi} \, \sigma} \exp \left[- \frac{1}{2}\left(\frac{w - \mu_1 - \mu_2}{\sigma}\right)^2 \right],
\end{align}
where $\sigma \deff  \sqrt{\sigma_1^2 + 2 \rho \, \sigma_1 \sigma_2 + \sigma_2^2}$. This is the univariate normal distribution $\mathcal{N}_1(\mu_1 + \mu_2, \sigma)$, so that using \eqref{e20ter} we find $\expe{W} = \mu_1 + \mu_2$, in agreement with \eqref{e90}.

Next, we use the second method of integration to calculate the characteristic function $\varphi_W(s)$. From \eqref{delta30} we have
\begin{align}\label{delta40}
\varphi_W(s) =  & \; \int_{\mathbb{R}^{2}}  e^{i s  \left( x_1 + x_2  \right) }  \, f_{X_1 X_2}(x_1 , x_2) \, \di x_1 \, \di x_2 \nonumber \\[4pt]
 =  & \;  \exp\left[ - \frac{1}{2} s^2 \sigma^2  - 2 \, i \, s \left( \mu_1 + \mu_2 \right) \right],
\end{align}
where, as before, $\sigma \deff  \sqrt{\sigma_1^2 + 2 \rho \, \sigma_1 \sigma_2 + \sigma_2^2}$. Substituting \eqref{delta40} into \eqref{delta20} we obtain again \eqref{e100}.

We will see later that the situation is radically different in quantum mechanics where the real random variables $X_1$ and $X_2$ will be replaced by the Hermitian operators $\hX_1$ and $\hX_2$.
In this case the sum $\hX_1 + \hX_2$ will be also an Hermitian operator. This implies  that the quantum analogous of $f_W(w)$ do exist, but the joint PDF $f_{X_1 X_2}(x_1,x_2)$ does not if $\hX_1 \hX_2 - \hX_2 \hX_1 \neq 0$.

\subsection{Product of two random variables}\label{ExampleProd}

\begin{figure}[ht!]
  \centering
  \includegraphics[scale=3,clip=false,width=1\columnwidth,trim = 0 0 0 0]{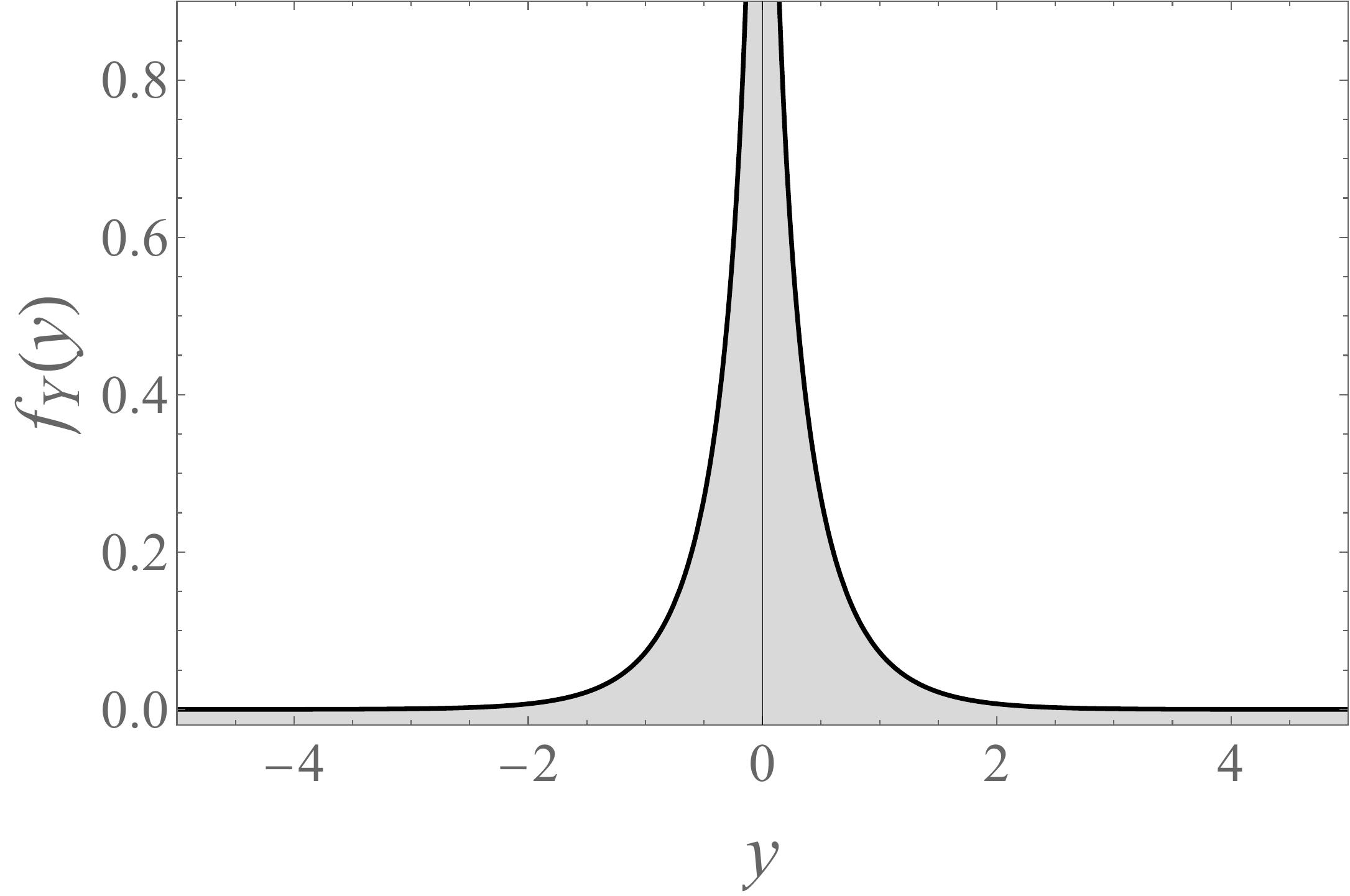}
  \caption{Plot of the probability density function $f_{Y}(y)$ given by Eq. \eqref{e60}. To generate this plot the we have used the following values: $\sigma_1 = 1/\sqrt{2} = \sigma_2$, and $\rho = 0$. At $y= 0$ the PDF $f_{Y}(y)$ has an integrable logarithmic singularity. When $\rho \neq 0$, the curve becomes non symmetric around $y=0$.}\label{fig1}
\end{figure}

For clarity, let us take $\mu_1 = 0 = \mu_2$ in \eqref{e20}, and  consider  the \emph{product} of  $X_1$ and $X_2$:   $g(X_1,X_2) = X_1 X_2$. From \eqref{a120} it follows that
\begin{align}\label{e30}
\expe{X_1 X_2} = & \;  \int_{\mathbb{R}^2}  x_1 x_2 \, f_{X_1 X_2}(x_1,x_2) \, \di x_1 \di x_2 \nonumber \\[4pt]
= & \; \rho \, \sigma_1 \sigma_2.
\end{align}
Now we can proceed exactly as in the previous example,  taking $Y = X_1  X_2$, with $Y \sim f_Y(y)$, to eventually calculate
\begin{align}\label{e35}
\expe{Y} = \int_{\mathbb{R}}  y \, f_Y(y) \, \di y.
\end{align}
The  first method of integration of \eqref{a192} with $x_2$ as first integration variable, gives
\begin{align}\label{e50}
f_{Y}(y)
= & \; \expe{\delta  \bigl(y - X_1 X_2   \bigr)} \nonumber \\[4pt]
= & \; \int_{\mathbb{R}}  \frac{1}{\abs{x_1}} \, f_{X_1 X_2} \left( x_1, \frac{y}{x_1} \right) \, \di x_1 .
\end{align}
This integral can be calculated exactly by Mathematica \cite{Mathematica}, which gives
\begin{align}\label{e60}
f_{Y}(y) = \frac{1}{\pi D} \exp \left({y \frac{\rho}{D}} \right) K_0\left( \frac{\abs{y}}{D}\right),
\end{align}
where $D\deff  \sqrt{\det V} = \sigma_1 \sigma_2 \sqrt{1 - \rho^2}$,
and $K_0(z)$ is the modified Bessel function of the second kind \cite{ModifiedBesselSec}.  A plot of this function is shown in Fig. \ref{fig1}.
%
%
%
%
%
%
 Notice that $f_{Y}(y)$ has an integrable logarithmic  singularity at $y=0$, so that
\begin{align}\label{e70}
\int_{\mathbb{R}} f_Y(y)\, \di y = 1.
\end{align}
Substituting \eqref{e60} into \eqref{e35} we obtain,
\begin{align}\label{e80}
\expe{Y} = \rho \, \sigma_1 \sigma_2,
\end{align}
in agreement with \eqref{e30}.

As in the previous example, the characteristic function of $Y$ can be calculated using \eqref{delta30}. The result is
\begin{align}\label{delta50}
\varphi_Y(s) =  & \;  \int_{\mathbb{R}^{2}}  e^{i s \,  x_1  x_2  }  \, f_{X_1 X_2}(x_1 , x_2) \, \di x_1 \, \di x_2 \nonumber \\[4pt]
 =  & \; \frac{1}{ \sqrt{1 + s^2 \left(1-\rho^2 \right) \sigma_1^2 \sigma_2^2 - 2 \, i \, s \, \rho \, \sigma_1 \sigma_2}} .
\end{align}
It should be noticed that in this case the first method of integration of \eqref{a192} is more convenient because if we substitute \eqref{delta50} into \eqref{delta20}, Mathematica is not able to perform the integral that would reproduce \eqref{e60}.

Thus, we have found that both the \emph{joint} PDF $f_X(x_1,x_2)$ for the \emph{two} random variables $X_1$ and $X_2$, and the  PDF $f_Y(y)$ for the \emph{single} random variable $Y = X_1 X_2$, do exist and are well defined, and  yields $\expe{X_1 X_2} = \expe{Y}$.
We will show later that in the corresponding quantum case the product $\hX_1 \hX_2$ will be an Hermitian operator corresponding to $Y = X_1 X_2$, only if $\hX_1 \hX_2 - \hX_2 \hX_1 =0$. When this condition fails, the quantum analogous of \emph{both}  PDFs  $f_Y(y)$ and $f_{X_1 X_2}(x_1,x_2)$ do not exist.

\section{The spectral theorem}\label{SpecQuant}

In both classical and quantum mechanics, physical systems are characterized by some quantities, the so-called physical observables (observables, for short), which can be actually measured, such as position, momentum, angular momentum and energy. In classical mechanics an observable is represented by a real number or a real vector, which provides the observed value of the quantity when a measurement is made on the system. In quantum mechanics, on the other hand, an observable is represented by a Hermitian operator that may or may not have a definite value, depending on the manner in which the system is prepared \cite{NoteHermitian}. The ``manner in which a system is prepared'' is called the \emph{quantum state} of the system, and it is represented by a  vector in a complex vector space known as  Hilbert space. When the quantum state is an eigenvector of the operator representing the observable, the latter has a definite value provided by the eigenvalue associated with that eigenvector. Vice versa, if the quantum state is an arbitrary vector which is not an eigenvector of the operator representing the observable, the latter has not a definite value. In this case, measurements of the observable made on a collection of identically prepared systems will yield different values of the observable, not all of them equal. As a result, the  measured mean value of the observable will give the so-called \emph{expectation value} of the operator  with respect to the  quantum state in which the system was prepared \cite{Sakurai}.

In the light of what we have learned in the previous section, it is natural to wonder whether it would be possible, given a Hermitian operator $\hX$ representing some observable of a quantum system prepared into some state, to find a corresponding real-valued random variable $X$ whose expected value coincides with the expectation value of $\hX$  in that state. It turns out that the answer to this question is positive, and it is provided by the \emph{spectral theorem} \cite{jordan1997,Ballentine}. The latter states that given a Hermitian operator $\hX$  there is a unique family of projection operators $\hE_X(\lambda)$ depending on the real parameter $\lambda$, the  so-called \emph{spectral family},  such that given  a normalized arbitrary vector $\ket{\psi}$, we have
\begin{align}\label{a200}
\mean{\psi}{\hX}{\psi} =  \int_{\mathbb{R}} \lambda\, f_X^{\psi}(\lambda) \, \di  \lambda,
\end{align}
where here and hereafter $\brak{\psi}{\psi}=1$, and
\begin{align}\label{a210}
f_{X}^{\psi}(\lambda) \, \di  \lambda\deff   \di  \mean{\psi}{\hE_X(\lambda)}{\psi}  ,
\end{align}
is the probability density function of the random variable $X \sim f_{X}^{\psi}(\lambda)$ associated with the operator $\hX$.
In the last two equations the subscript ``$X$'' and the superscript ``$\psi$''  remind that  $f_{X}^{\psi}(\lambda)$ is determined by \emph{both} the operator $\hX$ and the quantum state $\ket{\psi}$, because different quantum states produce distinct expectation values of the same operator. A rigorous proof of this theorem can be found in section 120 of \cite{Riesz_Nagy}.

From \eqref{a30} it follows that the  cumulative distribution function $F_X^{\psi}(\lambda)$ associated with  $f_{X}^{\psi}(\lambda)$, is given by $f_{X}^{\psi}(\lambda) \, \di \lambda= \di F_X^{\psi}(\lambda)$.   Then, \eqref{a210} implies
\begin{align}\label{a212}
F_X^{\psi}(\lambda) =   \mean{\psi}{\hE_X(\lambda)}{\psi}  .
\end{align}
Since any CDF $F_X(x)$ is a monotonic increasing function satisfying $\lim_{x \to - \infty} F_X(x) = 0$, and $\lim_{x \to  \infty} F_X(x) = 1$, we may infer   that a spectral family should obey the following boundary conditions:
\begin{equation}\label{a215}
\begin{split}
 \hE_X(\lambda) \to \hat{O}  \quad & \text{for} \quad \lambda\to - \infty,  \\[4pt]
 \hE_X(\lambda) \to \hat{I}  \quad & \text{for} \quad \lambda\to  \infty.
\end{split}
\end{equation}
where $\hat{O}$ and $\hI$ are the zero and the identity operators, respectively.

Our next goal is to find a simple expression for $\hE_X(\lambda)$  using only elementary math and the elegant Dirac formulation of quantum mechanics. For this purpose we will follow closely chapters $1$ and $2$ of Ballentine's book \cite{Ballentine}, and we will consider the cases of continuous and discrete spectra separately.

\section{Continuous spectrum}\label{continuous}

Let $\hX$ be a Hermitian operator having a purely continuous spectrum \cite{NoteSakurai}:
\begin{align}\label{a220}
\hX \ket{\lambda_X} = \lambda\ket{\lambda_X}, \qquad (\lambda\in \mathbb{R}).
\end{align}
The ket $\ket{\lambda_X}$ is an eigenvector of the operator $\hX$ with eigenvalue $\lambda$. These eigenvectors have the following properties:
\begin{align}
\brak{\lambda_X}{\lambda_X'} = \delta \left(\lambda- \lambda' \right), \qquad (\textrm{orthogonality}), \label{a230} \\[4pt]
\int_{\mathbb{R}} \proj{\lambda_X}{\lambda_X} \, \di \lambda= \hI, \qquad (\textrm{completeness}), \label{a240}
\end{align}
where $\hI$ denotes the identity operator.
Using \eqref{a220} and \eqref{a240}, we can rewrite the left-hand side of \eqref{a200}, as
\begin{align}\label{a250}
\mean{\psi}{\hX}{\psi} = & \;  \mean{\psi}{\hX \hI}{\psi} \nonumber \\[4pt]
= & \;  \int_{\mathbb{R}} \mean{\psi}{\hX}{\lambda_X}\brak{\lambda_X}{\psi} \, \di \lambda\nonumber \\[4pt]
= & \;  \int_{\mathbb{R}} \lambda\abs{\brak{\lambda_X}{\psi}}^2  \di \lambda.
\end{align}
Comparison between \eqref{a200} and \eqref{a250} gives
\begin{align}\label{a260}
f_{X}^{\psi}(\lambda) = \abs{\brak{\lambda_X}{\psi}}^2 .
\end{align}
The physical meaning of this formula should be clear: given a physical system prepared in the state $\ket{\psi}$, the probability $f_{X}^{\psi}(\lambda) \di \lambda$ that the operator $\hX$ takes a value in the interval $(\lambda,\lambda + \di \lambda]$ when a measurement of the observable represented by $\hX$ is performed upon the system, is equal to the probability that the corresponding random variable $X$ takes a value within the same interval.

The function $f_{X}^{\psi}(\lambda)$ is a proper PDF because $\abs{\brak{\lambda_X}{\psi}}^2 \geq 0$, and
\begin{align}\label{a270}
\int_{\mathbb{R}} \abs{\brak{\lambda_X}{\psi}}^2 \, \di \lambda= \brak{\psi}{\psi} = 1,
\end{align}
where \eqref{a240} has been used.
We are now able to rewrite \eqref{a210} as
\begin{align}\label{a280}
\abs{\brak{\lambda_X}{\psi}}^2 \, \di \lambda=   \di  \mean{\psi}{\hE_X(\lambda)}{\psi} .
\end{align}
Integrating both sides of \eqref{a280} for $-\infty < \lambda\leq x$, and using \eqref{a215}, we obtain
\begin{align}\label{a290}
\int_{-\infty}^x \abs{\brak{\lambda_X}{\psi}}^2  \di \lambda= & \; \int_{\mean{\psi}{\hE_X(-\infty)}{\psi}}^{\mean{\psi}{\hE_X(x)}{\psi}}\di  \mean{\psi}{\hE_X(\lambda)}{\psi} \nonumber \\[8pt]
= & \; \mean{\psi}{\hE_X(x)}{\psi}.
\end{align}
This equation can be rearranged as
\begin{align}\label{a300}
\mean{\psi}{\hE_X(x)}{\psi} = \bra{\psi} \left( \int_{-\infty}^x \proj{\lambda_X}{\lambda_X} \, \di \lambda\right) \ket{\psi}.
\end{align}
Since this must hold for arbitrary $\ket{\psi}$, we may infer that
\begin{align}\label{a310}
\hE_X(x) = & \;   \int_{-\infty}^x \proj{\lambda_X}{\lambda_X} \, \di \lambda, \nonumber \\[4pt]
= & \;   \int_{-\infty}^\infty \Hs ( x - \lambda) \proj{\lambda_X}{\lambda_X} \, \di \lambda\nonumber \\[4pt]
= & \;   \int_{-\infty}^\infty \Hs ( x - \hX ) \proj{\lambda_X}{\lambda_X} \, \di \lambda\nonumber \\[4pt]
= & \;   \Hs ( x - \hX ),
\end{align}
where we have used \eqref{a240}, and the trivial identity
\begin{align}\label{a320}
h( \hX ) \ket{\lambda_X} = h(\lambda) \ket{\lambda_X},
\end{align}
valid for any  function $h(x)$. We have thus determined the projection operators $\hE_X(x)$ forming the spectral family of the operators $\hX$.

The result \eqref{a310} is important because it allows us to make a direct connection between quantum operators and quantum states on the one hand, and random variables on the other. To see this,  we use \eqref{a310} to rewrite \eqref{a212} as
\begin{align}\label{a340}
F_{X}^{\psi}(x) & =    \mean{\psi}{\Hs ( x - \hX )}{\psi}  \nonumber \\[4pt]
& \deff   \ave{\Hs ( x - \hX )}_\psi ,
\end{align}
where the notation used in the second line is commonplace in textbooks on quantum mechanics (see, e.g., \cite{GalindoI}). This equation should be compared with \eqref{a80g} to appreciate the similarity between quantum and classical probability distributions:
\begin{align}\label{a342}
F_X(x) = \expe{\Hs (x - X)}. \tag{\ref{a80g}}
\end{align}

Next, taking the derivative of both sides of \eqref{a340} with respect to $x$, we find
\begin{align}\label{a350}
f_{X}^{\psi}(x) & =  \mean{\psi}{\delta ( x - \hX )}{\psi}  \nonumber \\[4pt]
& \deff   \ave{\delta ( x - \hX )}_\psi .
\end{align}
Again, this equation should be compared with \eqref{a80e}:
\begin{align}\label{a351}
 f_X(x) = \expe{\delta \left( x - X  \right)}  \, . \tag{\ref{a80e}}
\end{align}

Before discussing the implications of these results in the next section, let us notice that from \eqref{a200} and \eqref{a350}, it follows that given any real-valued function $g=g(x)$, such that  $g: \mathbb{R} \to \mathbb{R}$, the expectation value of the operator $g(\hX)$ with respect to the vector $\ket{\psi}$ is given by
\begin{align}\label{a352}
\mean{\psi}{g(\hX)}{\psi} =  \int_{\mathbb{R}} g(x) \, \mean{\psi}{\delta ( x - \hX )}{\psi} \, \di  x .
\end{align}
Since this must hold for arbitrary $\ket{\psi}$, then \eqref{a352}   implies
\begin{align}\label{a354}
g(\hX) =  \int_{\mathbb{R}} g(x) \, \delta ( x - \hX ) \, \di  x,
\end{align}
which is just another common way, but not as trivial as it may seem at first glance,  to express the spectral theorem (see, for example, section 15 of \cite{jordan1997}, for a formal proof). Equation \eqref{a352} is frequently used in quantum field theory but it is rarely found  in textbooks. An exception is the book presenting Sydney Coleman's wonderful lectures on quantum field theory \cite{Coleman2019b}.

\subsection{Two technical remarks}\label{technical}

The operator $\delta ( \lambda- \hX )$ is effectively a projector, as shown in  \cite{Xu2021}. This can be easily seen using \eqref{a240} and \eqref{a320}:
\begin{align}\label{a360}
\delta ( \lambda- \hX ) = & \;   \delta ( \lambda- \hX ) \hI  \nonumber \\[4pt]
= & \;   \int_{\mathbb{R}} \delta ( \lambda- \hX ) \proj{\lambda_X'}{\lambda_X'} \, \di \lambda'  \nonumber \\[4pt]
= & \;   \int_{\mathbb{R}} \delta ( \lambda- \lambda' ) \proj{\lambda_X'}{\lambda_X'} \, \di \lambda'  \nonumber \\[4pt]
= & \;   \proj{\lambda_X}{\lambda_X} \, .
\end{align}
Substituting this result in \eqref{a350}, we obtain again \eqref{a260}.

For all practical purposes, $\delta ( \lambda- \hX )$ is conveniently understood in terms of the delta distribution integral representation \eqref{a80h}, that is
\begin{align}\label{a370}
\delta ( \lambda- \hX ) =  \frac{1}{2 \pi} \int_{\mathbb{R}}  e^{-i s ( \lambda- \hX )} \,  \di s .
\end{align}
By analogy with the classical case \eqref{a80i}, we can use this representation to introduce the \emph{quantum characteristic function} $\varphi_X^\psi(s)$, defined via
\begin{align}\label{a372}
f_X^\psi(x)  & =\mean{\psi}{\delta(x - \hX)}{\psi} \nonumber \\[4pt]
 & = \frac{1}{2 \pi} \int_{\mathbb{R}} e^{-i s x }  \mean{\psi}{ e^{i s \hX }}{\psi}  \,  \di s \nonumber \\[4pt]
 & \deff  \frac{1}{2 \pi} \int_{\mathbb{R}} e^{-i s x }  \varphi_X^\psi(s)  \,  \di s.
\end{align}
This equation is the quantum counterpart of Eq. \eqref{a83}.

\section{Discrete spectrum}\label{discrete}

When one goes from continuous spectra to discrete ones, things go very smoothly.
Let $\hA$ be a Hermitian operator having a purely discrete spectrum \cite{NoteSakurai}:
\begin{align}\label{f10}
\hA \ket{a_n} = a_n \ket{a_n}, \qquad (n \in \mathbb{N}).
\end{align}
The ket $\ket{a_n}$ is an eigenvector of the operator $\hA$ with eigenvalue $a_n$. These eigenvectors have the following properties:
\begin{align}
\brak{a_n}{a_m} = \delta_{nm},  \qquad (\textrm{orthogonality}), \label{f20} \\[4pt]
\sum_{n \in \mathbb{N}} \proj{a_n}{a_n} = \hI, \qquad (\textrm{completeness}), \label{f30}
\end{align}
where $n,m \in \mathbb{N}$.

Let $A$ be a random variable associated with the quantum operator $\hA$ via the spectral theorem, with CDF $F_A^\psi(\lambda)$, and PDF $f_A^\psi(\lambda)$.   We leave it to the reader as a simple exercise, to show that given the quantum state vector
\begin{align}\label{f40}
\ket{\psi} = \sum_{n \in \mathbb{N}} \ket{a_n} \brak{a_n}{\psi},
\end{align}
one can rewrite the projector $\hE_A(\lambda)$ from \eqref{a310}, as
\begin{align}\label{f90}
\hE_A(\lambda) = & \; \Hs(\lambda - \hA) \nonumber \\[4pt]
 = & \; \sum_{n \in \mathbb{N}} \Hs(\lambda - \hA) \proj{a_n}{a_n} \nonumber \\[4pt]
 = & \; \sum_{n \in \mathbb{N}} \Hs(\lambda - a_n) \proj{a_n}{a_n},
\end{align}
where the completeness relation \eqref{f30} has been used. From this equation and \eqref{a340} it follows that
\begin{align}\label{f50}
F_A^\psi(\lambda) = & \; \mean{\psi}{\hE_A(\lambda)}{\psi} \nonumber \\[4pt]
= & \; \sum_{n \in \mathbb{N}} p_A^\psi(\lambda) \, \Hs(\lambda - a_n),
\end{align}
where $p_A^\psi(\lambda)$ is the probability mass function of $A$ defined by
\begin{align}\label{f60}
p_A^\psi(\lambda) = \prob{A = \lambda},
\end{align}
with
\begin{align}\label{f70}
p_A^\psi(a_n) = \abs{\brak{a_n}{\psi}}^2.
\end{align}
Tacking the derivative of \eqref{f50} with respect to $\lambda$, one obtains the PDF
\begin{align}\label{f80}
f_A^\psi(\lambda) = \sum_{n \in \mathbb{N}} p_A^\psi(\lambda) \, \delta(\lambda - a_n).
\end{align}
Finally, substituting \eqref{f80} into \eqref{a200}, we obtain
\begin{align}\label{f100}
\mean{\psi}{\hA}{\psi} = \sum_{n \in \mathbb{N}} a_n \, p_A^\psi(a_n).
\end{align}
The similarity between the ``quantum'' expression \eqref{f80}, and the ``classical'' one \eqref{a54}, reproduced below, should be noticed:
\begin{align}\label{f110}
f_X(x) = \sum_{m=1}^{M} p_X(x)  \delta ( x - x_m ). \tag{\ref{a54}}
\end{align}

\section{Random variables in classical and quantum mechanics}\label{first}

We have just seen how an informal use of generalized functions and Dirac's ket and bra notation, makes it possible to obtain simple expressions for quantum probability distributions that are formally identical to the classical ones.
In practice, as far as we are concerned with a \emph{single} observable of a physical system prepared in a specific manner described by the quantum state vector $\ket{\psi}$, the spectral theorem gives us a one-to-one correspondence between the Hermitian quantum operator $\hX$ representing the observable, and the real-valued random variable $X \sim f_X^\psi(x) = \ave{\delta ( x - \hX )}_\psi$. The importance of the spectral theorem lies in the fact that it not only gives us the correspondence between $\hX$ and $X$, but also the explicit recipe for building the PDF of $X$.

Unfortunately, things are not always so pleasant when we consider the case of two or more observables of a quantum system. As we shall see, in this case there is a striking difference between compatible and incompatible observables.

\subsection{Compatible observables}\label{compatible}

In quantum mechanics, \emph{compatible observables} are represented by Hermitian operators that commute with each other \cite{Sakurai}. In this case the generalization of the spectral theorem to two or more Hermitian operators is straightforward (see, for example, sections 16 and 25 of \cite{jordan1997} for a technical discussion).
Let $\{ \hX_1, \ldots, \hX_N \}$ be  a complete set of $N$ commuting operators, and let $\ket{\psi}$ be a normalized vector. Then, the \emph{joint} PDF $f_{X_1 \cdots X_N}^\psi(x_1, \ldots,x_N) = f_{X_1 \cdots X_N}^\psi(\bx)$ of the random vector $\bX = (X_1, \ldots, X_N)$, such that
\begin{align}\label{a2000}
\mean{\psi}{\hX_n}{\psi} =  \int_{\mathbb{R}^N} x_n \, f_{X_1 \cdots X_N}^\psi(\bx) \, \di^N x,
\end{align}
where $\di^N x \deff  \di x_1 \ldots \di x_N$, can be found simply by rewriting \eqref{a160}-\eqref{a190}  with the random variables $(X_1, \ldots, X_N)$ replaced by the corresponding operators $(\hX_1, \ldots, \hX_N)$:
\begin{align}\label{d120}
f_{X_1 \cdots X_N}^\psi(x_1, \ldots,x_N) =  \mean{\psi}{\prod_{n=1}^N \delta \bigl( x_n - \hX_n  \bigr)}{\psi}.
\end{align}
It should be noticed that since the operators $\{ \hX_1, \ldots, \hX_N \}$ commute with each other, the product above is perfectly well-defined and any choice of the order of the $N$ delta distributions is permitted.
The demonstration of this statement amounts to a straightforward calculation. Let $g(x_1, \ldots, x_n)\deff g(\bx)$ be a function $g: \mathbb{R}^N \to \mathbb{R}$ such that $[g(\hX_1, \ldots, \hX_n)]^\dagger = g(\hX_1, \ldots, \hX_N)\deff  g(\hat{\bX})$, and let $\{ \ket{ x_1', \ldots, x_N'} \}\deff  \{ \ket{ \bx'} \}$ be a complete orthonormal basis for the Hilbert space of the system, such that $\hX_n \ket{ \bx'} = x_n' \ket{ \bx'}$. Then, using the completeness relation
\begin{align}\label{d125}
 \int_{\mathbb{R}^N} \proj{\bx'}{\bx'} \, \di^N x'  = \hI,
\end{align}
it is not difficult to show that
\begin{align}\label{d130}
\mean{\psi}{g(\hat{\bX})}{\psi} = & \; \int_{\mathbb{R}^N} \mean{\psi}{g(\hat{\bX})}{\bx'} \brak{\bx'}{\psi}\, \di^N x' \nonumber \\[4pt]
= & \; \int_{\mathbb{R}^N} g(\bx') \abs{\brak{\bx'}{\psi}}^2 \, \di^N x'.
\end{align}
This equation implies
\begin{align}\label{d140}
f_{X_1 \cdots X_N}^\psi(\bx) = \abs{\brak{\bx}{\psi}}^2 ,
\end{align}
which is just the $N$-dimensional  generalization of $\eqref{a260}$. This result is clearly in agreement with \eqref{d120} because
\begin{align}\label{d150}
f_{X_1 \cdots X_N}^\psi(\bx) & = \bra{\psi}\prod_{n=1}^N  \delta \bigl( x_n - \hX_n  \bigr) \ket{\psi} \nonumber \\[4pt]
& =  \int_{\mathbb{R}^N}  \mean{\psi}{\prod_{n=1}^N \delta \bigl( x_n - \hX_n  \bigr)}{\bx'} \brak{\bx'}{\psi} \nonumber  \\[4pt]
& =  \int_{\mathbb{R}^N}  \prod_{n=1}^N \delta \bigl( x_n - x_n'  \bigr)\abs{\brak{\bx'}{\psi}}^2  \di^N x' \nonumber
\\[4pt]
& = \abs{\brak{\bx}{\psi}}^2 .
\end{align}

\subsubsection{Examples}\label{TwoExamples}

Consider the three orthogonal components of the vector position operator $\hat{\bm{Q}} = (\hQ_1, \hQ_2, \hQ_3)$. These operators commute with each other:  $\bigl[\hQ_n,\hQ_m \bigr]=0$, and have a continuous spectrum: $\hQ_n \ket{q_1,q_2,q_3} = q_n \ket{q_1,q_2,q_3}, ~ q_n \in \mathbb{R}$, with $(n,m=1,2,3)$. In this case \eqref{d150} gives the joint PDF $f^\psi_{Q_1 Q_2 Q_3}(q_1,q_2,q_3)$, which is defined by
\begin{align}\label{d77}
f^\psi_{Q_1 Q_2 Q_3}(q_1,q_2,q_3) = \abs{\brak{q_1,q_2,q_3}{\psi}}^2.
\end{align}

For operators with discrete spectra, things are the same way. Let  $\hA$ and $\hB$ be commuting operators with discrete eigenvalues $a_n$ and $b_m$ associated with the common eigenvector $\ket{a_n,b_m}$. Born's rule tells us that given the quantum state vector $\ket{\psi}$, the probability that operator $\hA$ takes the value $a_n$, and operator $\hB$ takes the value $b_m$, is simply given by $\abs{\brak{a_n, b_m}{\psi}}^2$. This implies the existence of the joint probability mass function
\begin{align}\label{d70}
p_{XY}^\psi(a_n,b_m) = \abs{\brak{a_n, b_m}{\psi}}^2,
\end{align}
and of the joint PDF
\begin{align}\label{d75}
f_{XY}^\psi(x,y) = \sum_{n,m \in \mathbb{N} }  p_{XY}^\psi(a_n,b_m) \delta(x-a_n) \delta(y-b_n).
\end{align}

\subsection{Incompatible observables}\label{incompatible}

Things are getting more complicated when dealing with two or more \emph{incompatible observables}, which are represented by operators that do not commute  with each other \cite{Sakurai}. To show this, consider two observables represented by the Hermitian operators $\hX$ and $\hY$ such that $\bigl[ \hX, \hY \bigr] = \hC \neq \hO$. From the spectral theorem \eqref{a200} we know that given a quantum state vector $\ket{\psi}$, there are two  random variables $X$ and $Y$, such that
\begin{equation}\label{d62}
\begin{split}
X \sim & \; f_{X}^\psi(x) = \mean{\psi}{\delta ( x - \hX )}{\psi}, \\[4pt]
Y \sim &\; f_{Y}^\psi(y) = \mean{\psi}{\delta ( y - \hY )}{\psi}.
\end{split}
\end{equation}
 Suppose the two observables are physically related to each other (think, for example, of the position and momentum of a particle at a given instant of time).
Then it would seem reasonable to assume the existence of a joint PDF, say $f_{X Y}^\psi(x,y)$, such that $\bX = (X,Y) \sim f_{X Y}^\psi(x,y)$. Such a PDF should have the standard properties of a probability density function, namely $f_{X Y}^\psi(x,y) \geq 0$, and
\begin{align}\label{d60}
\int_{\mathbb{R}^2} f_{X Y}^\psi(x,y) \, \di x \, \di y  = 1.
\end{align}
Moreover, it should generate the correct marginal density functions $f_{X}^\psi(x)$ and $f_{Y}^\psi(y)$, which are known from \eqref{d62}:
\begin{equation}\label{d40}
\begin{split}
f_{X}^\psi(x) = & \; \int_{\mathbb{R}} f_{X Y}^\psi(x,y) \, \di y = \mean{\psi}{\delta ( x - \hX )}{\psi}, \\[4pt]
f_{Y}^\psi(y) = & \; \int_{\mathbb{R}} f_{X Y}^\psi(x,y) \, \di x = \mean{\psi}{\delta ( y - \hY )}{\psi}.
\end{split}
\end{equation}
However, it is not difficult to show that such $f_{X Y}^\psi(x,y)$ cannot exist.

For this purpose, let us suppose  that the two random variables $X$ and $Y$ defined by \eqref{d62} could be \emph{sampled simultaneously} \cite{papoulis2002}. This means that there would be an experiment (remember our definition of ``experiment'' in Sec. \ref{oneRV}) yielding
a \emph{single} outcome $\omega \in \Omega$ such that $X(\omega) = x$ and  $Y(\omega) = y$, where $\Omega$ denotes, as usual, the sample space of the experiment. This is a key point for simultaneous sampling: a single outcome of the experiment determines the value of \emph{both} random variables.
 By definition, the probability that in a \emph{single run} of the experiment the outcome $\omega$  occurs, so that $X$ would take a value in the interval $(x, x + \di x]$, and simultaneously $Y$ would take a value in the interval $(y, y + \di y]$, is simply given by the joint probability density function $f_{XY}^\psi(x,y)$  \cite{papoulis2002}:
\begin{multline}\label{c10}
f_{XY}^\psi(x,y) \, \di x \, \di y \\[4pt]
 = \text{prob} \bigl\{x < X \leq x + \di x, y < Y \leq y + \di y \bigr\},
\end{multline}
for all $x,y \in \mathbb{R}$. However, if $X$ and $Y$  could be sampled simultaneously so that $X(\omega) = x$ and $Y(\omega) = y$, then the corresponding operators $\hX$ and $\hY$ could take the   values $x$ and $y$ simultaneously.
This means that there would exist a hypothetical orthonormal basis $\{ \ket{x_X, y_Y} \}$ of simultaneous eigenvectors of $\hX$ and $\hY$,  such that $\hX \ket{x_X, y_Y} = x \ket{x_X, y_Y}$ and $\hY \ket{x_X, y_Y} = y \ket{x_X, y_Y}$ for all $x,y \in \mathbb{R}$.
However, it is not difficult to show that if such a basis exists, then $\hX$ and $\hY$ must commute.
To this end, suppose  that the hypothetical basis vectors $\ket{x_X, y_Y}$ above do exist. This would imply
\begin{align}\label{ab10}
\bigl[ \hX, \hY \bigr] \ket{x_X, y_Y} = & \; \hX \bigl( \hY \ket{x_X, y_Y} \bigr)  - \hY \bigl( \hX \ket{x_X, y_Y} \bigr) \nonumber \\[4pt]
= & \; \left( y \, x - x \, y \right) \ket{x_X, y_Y} \nonumber \\[4pt]
= & \; 0,
\end{align}
for all $x, y \in \mathbb{R}$. Next, consider the \emph{perfectly arbitrary} vector $\ket{\psi}: \brak{\psi}{\psi} =1$, defined by
\begin{align}\label{ab20}
 \ket{\psi} = \int_{\mathbb{R}^2} \psi(x, y) \ket{x_X, y_Y} \, \di x \, \di y,
\end{align}
where $\psi(x, y)$ is any square integrable complex-valued function: $\psi : \mathbb{R}^2 \to \mathbb{C}$.
Then, using \eqref{ab10} we could calculate straightforwardly
\begin{align}\label{ab30}
\bigl[ \hX, \hY \bigr] \ket{\psi} = & \; \int_{\mathbb{R}^2} \psi(x, y) \bigl[ \hX, \hY \bigr] \ket{x_X, y_Y} \, \di x \, \di y \nonumber \\[4pt]
= & \; 0.
\end{align}
Since this  must hold  for any vector $\ket{\psi}$, we may infer that $\bigl[ \hX, \hY \bigr]  = \hO$.  Therefore, such an orthonormal basis $\{ \ket{x_X, y_Y} \}$ does not exist if $\bigl[ \hX, \hY \bigr]  \neq \hO$.

So, we have learned that it is impossible to find a joint PDF for two  random variables corresponding to two noncommuting operators. However, we will see that it is still possible to find the so-called   quasi-probability distributions for such random variables, which are function satisfying \eqref{d60} and \eqref{d40} but are not everywhere nonnegative.
A very famous example of a  quasi-probability distribution is the Wigner function  (see, for example, \cite{Case2008} and references therein). In the next section we will show how  quasi-probability distributions arise in the case of random variables associated with the sum and product of two noncommuting operators. The corresponding classical cases were presented in Sec. \ref{Two_examples}.

\section{A case study: Sum and product of two noncommuting operators}\label{SumProduct}

To clarify the problems occurring with incompatible observables, we consider now  the sum and the product of two noncommuting Hermitian operators. The analog classical cases were illustrated in Sec. \ref{Two_examples}, where we had found that given the random vector $\bX = (X,Y) \sim f_{XY}(x,y)$ with $X \sim f_{X}(x)$ and $Y \sim f_{Y}(y)$, one could calculate the expected values $\expe{X+Y}$ and $\expe{XY}$ as
\begin{equation}\label{c20}
\begin{split}
\expe{X+Y} = & \; \int_{\mathbb{R}^2} (x + y)  f_{XY}(x,y) \, \di x \, \di y, \\[4pt]
= & \; \int_{\mathbb{R}} x  f_{X}(x) \, \di x  + \int_{\mathbb{R}} y  f_{Y}(y) \, \di y \\[4pt]
\expe{XY} =  & \; \int_{\mathbb{R}^2} x \,  y \,  f_{XY}(x,y) \, \di x \, \di y,
\end{split}
\end{equation}
where $f_{X}(x)$ and $f_{Y}(y)$ are the marginal PDF of $X$ and $Y$.

Now we will try to extend these results to the quantum realm,  showing that given  two Hermitian operators $\hX$ and $\hY$, such that $\bigl[ \hX, \hY \bigr]  \neq \hO$, and given a vector $\ket{\psi}$,  the quantum analogous of \eqref{c20} can be written in a similar manner as
\begin{equation}\label{c30}
\begin{split}
\ave{\hX + \hY}_\psi = & \; \int_{\mathbb{R}^2} (\lambda + \mu)  f(\lambda,\mu) \, \di \lambda \, \di \mu, \\[4pt]
= & \; \int_{\mathbb{R}} \lambda \,  f^\psi_X(\lambda) \, \di \lambda + \int_{\mathbb{R}} \mu  \, f^\psi_Y(\mu) \, \di \mu, \\[4pt]
\frac{1}{2}\ave{\hX \hY + \hY \hX}_\psi =  & \; \int_{\mathbb{R}^2} \lambda  \,  \mu  \,   f(\lambda,\mu) \, \di \lambda \, \di \mu,
\end{split}
\end{equation}
where $f^\psi_X(\lambda)$ and $f^\psi_Y(\mu)$ are defined by \eqref{d62}.
However, differently from \eqref{c20}, we will find that  $f(\lambda,\mu) \in \mathbb{R}$  is \emph{not}  identifiable  with a PDF because it can take negative values.

Here and below we assume that  $\hX$ and $\hY$ with $\bigl[ \hX, \hY \bigr] = \hat{C} \neq \hO$, satisfy the eigenvalue equations
\begin{align}\label{d210}
\hX \ket{\lambda_X} =   \lambda\ket{\lambda_X}, \quad  \hY \ket{\mu_Y} =   \mu \ket{\mu_Y} , \quad(\lambda, \mu \in \mathbb{R}),
\end{align}
where the eigenvectors are orthogonal: $\brak{\lambda_X}{\lambda_X'} = \delta(\lambda-\lambda')$ and $\brak{\mu_Y}{\mu_Y'} = \delta(\mu-\mu')$, and form complete bases for the Hilbert space of the system:
$\hI = \hI_X \deff  \int_\mathbb{R} \proj{\lambda_X}{\lambda_X} \, \di \lambda$ and $\hI=\hI_Y \deff  \int_\mathbb{R} \proj{\mu_Y}{\mu_Y} \, \di \mu$.

\subsection{Sum  of two noncommuting operators}\label{Sum}

Let us calculate
\begin{align}\label{c40}
\mean{\psi}{\hX + \hY}{\psi}  = & \; \mean{\psi}{\hX \hI_X  + \hY \hI_Y}{\psi} \nonumber \\[4pt]
 = & \;  \int_{\mathbb{R}} \lambda \, \brak{\psi}{\lambda_X} \brak{\lambda_X}{\psi} \, \di \lambda \nonumber \\[4pt]
 & + \int_{\mathbb{R}} \mu \, \brak{\psi}{\mu_Y} \brak{\mu_Y}{\psi} \, \di \mu \nonumber \\[4pt]
= & \; \int_{\mathbb{R}} \lambda \,  f^\psi_X(\lambda) \, \di \lambda + \int_{\mathbb{R}} \mu  \, f^\psi_Y(\mu) \, \di \mu.
\end{align}
This proves the second line of the first of equations \eqref{c30}.
This is a pretty standard calculation, but not a very enlightening one. So now let us try to rewrite this formula as a double integral using the identity operator again:
\begin{align}\label{c42}
\mean{\psi}{\hX + \hY}{\psi}  = & \;  \int_{\mathbb{R}} \lambda \, \brak{\psi}{\lambda_X} \mean{\lambda_X}{\hI_Y}{\psi} \, \di \lambda \nonumber \\[4pt]
 & + \int_{\mathbb{R}} \mu \, \mean{\psi}{\hI_X}{\mu_Y} \brak{\mu_Y}{\psi} \, \di \mu \nonumber \\[4pt]
 = & \;   \int_{\mathbb{R}^2} (\lambda + \mu ) z(\lambda,\mu) \, \di \lambda \, \di \mu ,
\end{align}
where we have defined
\begin{align}\label{d230}
z(\lambda,\mu) : = \brak{\psi}{\lambda_X} \brak{\lambda_X}{\mu_Y} \brak{\mu_Y}{\psi}.
\end{align}
In passing we note that this function can be also rewritten as follows:
\begin{align}\label{c45}
\bra{\psi}\delta(\lambda - \hX) & \delta(\mu - \hY) \ket{\psi}  \nonumber \\[4pt]
\; = & \;  \mean{\psi}{\delta(\lambda - \hX) \hI_X \delta(\mu - \hY) \hI_Y }{\psi} \nonumber \\[4pt]
\;  = & \;   \int_{\mathbb{R}^2} \delta(\lambda - \lambda')  \delta(\mu - \mu')   \nonumber \\[4pt]
\;  & \times \brak{\psi}{\lambda_X'} \brak{\lambda_X'}{\mu_Y'} \brak{\mu_Y'}{\psi} \, \di \lambda' \, \di \mu' \nonumber \\[4pt]
\;  = & \;   z(\lambda,\mu).
\end{align}
This  equation is reminiscent of the expression \eqref{d120} for the joint PDF of compatible observables. However, the connection is only formal because in this case $\delta(\lambda - \hX)  \delta(\mu - \hY)$ is not an Hermitian operator.

From $\mean{\psi}{\hX + \hY}{\psi} = \mean{\psi}{\hX + \hY}{\psi}^*$ if follows that we are permitted to rewrite
\begin{align}\label{c50}
\mean{\psi}{\hX + \hY}{\psi}  =   \int_{\mathbb{R}^2} (\lambda + \mu ) f(\lambda,\mu) \, \di \lambda \, \di \mu ,
\end{align}
where
\begin{align}\label{c60}
f(\lambda,\mu)  = \frac{z(\lambda,\mu) + z^*(\lambda,\mu)}{2} \in \mathbb{R}.
\end{align}
This proves the first line of the first of equations \eqref{c30}.

Via a simple direct calculation we can verify that
\begin{align}\label{c70}
\int_{\mathbb{R}^2} f(\lambda,\mu) \, \di \lambda \, \di \mu =1,
\end{align}
and that
\begin{equation}\label{c80}
\begin{split}
f_X^\psi(\lambda) = & \; \int_{\mathbb{R}} f(\lambda,\mu) \,  \di \mu = \abs{\brak{\lambda_X}{\psi}}^2, \\[4pt]
f_Y^\psi(\mu) =  & \; \int_{\mathbb{R}} f(\lambda,\mu) \, \di \lambda = \abs{\brak{\mu_Y}{\psi}}^2.
\end{split}
\end{equation}
Despite these appealing properties, $f(\lambda,\mu)$ cannot be regarded as a PDF because it may take negative values. Therefore, $f(\lambda,\mu)$ is an example of what  is typically called a quasi-probability distribution.

Next, by analogy with the ``classical'' equation \eqref{e40}, we want to show an alternative way to calculate $\mean{\psi}{\hX + \hY}{\psi}$. This method is based on  considering the Hermitian operator $\hat{W} = \hX + \hY $ as a whole, and  applying \eqref{a352}. In this way we find
\begin{align}\label{c90}
\mean{\psi}{\hW}{\psi}  = & \; \int_{\mathbb{R}} w \, \mean{\psi}{\delta(w - \hW)}{\psi} \, \di w  \nonumber \\[4pt]
 = & \;  \int_{\mathbb{R}} w \underbrace{\left[ \frac{1}{2 \pi} \int_{\mathbb{R}} e^{- i s w} \underbrace{\mean{\psi}{e^{i s (\hX + \hY)}}{\psi} }_{ = \;  \varphi_W^\psi(s)} \, \di s \right] }_{= \; f_W^\psi(w) }  \di w  \nonumber \\[4pt]
 = & \;  \int_{\mathbb{R}} w f_W^\psi(w)\, \di w,
\end{align}
where \eqref{a370} and \eqref{a372} have been used. From \eqref{a260} we know that if $\hW$ satisfy the eigenvalue equation $\hW \ket{\lambda_W} = \lambda \ket{\lambda_W}$, with $\lambda \in \mathbb{R}$, then
\begin{align}\label{c100}
f_W^\psi(\lambda) = \abs{\brak{\lambda_W}{\psi}}^2.
\end{align}

\subsection{Product of two noncommuting operators}\label{Product}

Let us consider now the product $\hX \hY$ and  calculate
\begin{align}\label{d212}
\mean{\psi}{\hX \hY}{\psi}  & = \mean{\psi}{\hX \hI_X \hY \hI_Y}{\psi} \nonumber \\[4pt]
 & = \int_{\mathbb{R}^2} \lambda \, \mu \, \brak{\psi}{\lambda_X} \brak{\lambda_X}{\mu_Y} \brak{\mu_Y}{\psi} \, \di \lambda \, \di \mu \nonumber \\[4pt]
 & \deff   \int_{\mathbb{R}^2} \lambda \, \mu \, z(\lambda,\mu) \, \di \lambda \, \di \mu ,
\end{align}
where $z(\lambda,\mu)$ is still defined by \eqref{d230}. This shows that $z(\lambda,\mu)$ is necessarily complex-valued because from  $\hX \hY \neq \hY \hX = (\hX \hY)^\dagger$ and $\mean{\psi}{\hY \hX}{\psi}  = \mean{\psi}{\hX \hY}{\psi}^*$, it follows that $z(\lambda,\mu)  \neq z^*(\lambda,\mu)$.
Therefore, since the product $\hX \hY$ is not a Hermitian operator, it cannot itself correspond to any physical quantity. However, it is always possible to rewrite $\hX \hY$  as
\begin{align}\label{d200}
\hX \hY = & \; \frac{\hX \hY + \hY \hX}{2} + i \, \frac{\hX \hY - \hY \hX}{2 i}\nonumber \\[4pt]
\deff  & \; \hU + i \, \hV,
\end{align}
where $\hU$ and $\hV$ are both Hermitian operators. The operator $\hU$ is the quantity commonly used to define the so-called \emph{quantum covariance} $\text{cov}(\hX,\hY)_\psi$ of the operators $\hX$ and $\hY$, which is defined by \cite{Levy-Leblond1986,Campos1998},
\begin{align}\label{g10}
\text{cov}(\hX,\hY)_\psi\deff  \frac{1}{2} \mean{\psi}{\hX \hY + \hY \hX}{\psi} - \mean{\psi}{\hX }{\psi} \mean{\psi}{\hY }{\psi}.
\end{align}
Using \eqref{d212} and $\mean{\psi}{\hY \hX}{\psi}  = \mean{\psi}{\hX \hY}{\psi}^*$, one can easily show that
\begin{align}\label{d220}
\ave{\hU}_\psi = & \; \frac{1}{2} \left( \mean{\psi}{\hX \hY}{\psi} +  \mean{\psi}{\hX \hY}{\psi}^* \right) \nonumber \\[4pt]
 = & \; \int_{\mathbb{R}^2} \lambda \, \mu \, f(\lambda,\mu) \, \di \lambda \, \di \mu .
\end{align}
This proves the second of equations \eqref{c30}. We remark that the function $f(\lambda,\mu)$ in the equation above is the \emph{same} as that appearing in equation \eqref{c50}.

Using again \eqref{d212}, we can also calculate
\begin{align}\label{d225}
\ave{\hV}_\psi = & \; \frac{1}{2 i} \left( \mean{\psi}{\hX \hY}{\psi} -  \mean{\psi}{\hX \hY}{\psi}^* \right) \nonumber \\[4pt]
 = & \; \int_{\mathbb{R}^2} \lambda \, \mu \, g(\lambda,\mu) \, \di \lambda \, \di \mu ,
\end{align}
where
\begin{align}\label{c60bis}
g(\lambda,\mu) \deff  \frac{z(\lambda,\mu) - z^*(\lambda,\mu)}{2 \, i } \in \mathbb{R}.
\end{align}
Note that although \eqref{d225} holds true,  $g(\lambda,\mu)$ is \emph{not} a quasi-probability distribution. In fact, a direct calculation shows that $g(\lambda,\mu)$ is not correctly normalized:
\begin{align}\label{d240}
\int_{\mathbb{R}} g(\lambda,\mu) \, \di \lambda \, \di \mu   = 0 .
\end{align}

As in the case of the sum of two operators, also $\hU$ and $\hV$ can be seen as operators in their entirety, by analogy with the classical equation \eqref{e50}. From the spectral theorem \eqref{a200} and \eqref{a350}, it follows that  the random variables $U$ and $V$ associated with the quantum operators  $\hU$ and $\hV$, will follow the probability density functions $f_{U}^\psi(u)$ and $f_{V}^\psi(v)$, respectively, defined by
\begin{align}
U \sim & \; f_{U}^\psi(u) = \mean{\psi}{\delta \left( u - \frac{\hX \hY + \hY \hX}{2} \right)}{\psi}, \label{d311} \\[4pt]
V \sim &\; f_{V}^\psi(v) = \mean{\psi}{\delta \left( v - \frac{\hX \hY - \hY \hX}{2 i} \right)}{\psi}, \label{d312}
\end{align}
with $u,v \in \mathbb{R}$. This will permit us to calculate again $\ave{\hU}_\psi$ and  $\ave{\hV}_\psi$, but this time  as
\begin{align}
\ave{\hU}_\psi = & \; \int_\mathbb{R} u \, f_{U}^\psi(u) \, \di u, \label{d220bis} \\[4pt]
\ave{\hV}_\psi = &\; \int_\mathbb{R} v \, f_{V}^\psi(v) \, \di v . \label{d225bis}
\end{align}

Our next task is to calculate explicitly the expectation value $\ave{\hX + \hY}_\psi$ using both \eqref{c50} and  \eqref{c90} for a specific choice of the operators $\hX$ and $\hY$ and of the vector $\ket{\psi}$. Then, we will also calculate concretely $\ave{\hU}_\psi$ and  $\ave{\hV}_\psi$ applying both (\ref{d220}-\ref{d225})  and (\ref{d220bis}-\ref{d225bis}) for the same choice of  $\hX$ and $\hY$ and  $\ket{\psi}$. These calculations will tangibly illustrate the concepts presented in this section.

\section{Two  examples again}\label{Examples}

In this section we provide  a practical application of the formulas given in the previous section.
Specifically, we  take the position $\hQ$ and momentum $\hP$ operators of a harmonic oscillator as an example of noncommuting Hermitian operators with continuous spectrum (see, for example, Sec. 2.3 of Ref. \cite{Sakurai} for a detailed treatment of the harmonic oscillator).

In quantum mechanics a harmonic oscillator with frequency $\omega$ and mass $m$ is described by the Hamilton operator
\begin{align}\label{d262}
\hH  =   \frac{1}{2m} \hP^2 + \frac{m \omega^2}{2} \, \hQ^2,
\end{align}
where $\hQ$ and $\hP$ are the position and momentum operators, respectively. This  Hamiltonian possesses a discrete spectrum:
\begin{align}\label{d263}
\hH \ket{n} = E_n \ket{n}, \qquad (n =0,1,2, \ldots),
\end{align}
where $E_n=(n + 1/2) \hbar \omega$.  The eigenvectors $\ket{n}$ form an orthonormal and complete basis of the Hilbert space of the harmonic oscillator:
\begin{align}
\brak{n'}{n} = \delta_{n' n},  \qquad (\textrm{orthogonality}), \label{f20bis} \\[4pt]
\sum_{n =0}^\infty \proj{n}{n} = \hI, \qquad (\textrm{completeness}), \label{f30bis}
\end{align}
where $n,n' = 0,1,2, \ldots$.

The position and momentum operators do not commute:
\begin{align}\label{d268}
\bigl[\hQ, \hP \bigr] = i \hbar ,
\end{align}
and exhibit a continuous spectrum:
\begin{align}\label{d270}
\hQ \ket{q_Q} =   q \ket{q_Q}, \quad  \hP \ket{p_P} =   p \ket{p_P} ,
\end{align}
with $q,p \in \mathbb{R}$.
Moreover, the eigenvectors $\ket{q_Q}$ and $\ket{p_P}$ are orthogonal and complete in the sense of \eqref{a230} and \eqref{a240}, but are not orthogonal to each other:
\begin{align}\label{d275}
\brak{q_Q}{p_P} =  \frac{1}{\sqrt{2 \pi \hbar}} \exp \left(i \, \frac{q \, p}{\hbar} \right).
\end{align}

For many applications it is convenient to introduce the  annihilation and creation operators $\ha$ and $\had$, respectively, defined by
\begin{align}\label{d380}
\ha \deff  \frac{1}{\sqrt{2}} \left(\frac{\hQ}{q_0} + i \frac{\hP}{p_0} \right), \quad \had \deff  \frac{1}{\sqrt{2}} \left(\frac{\hQ}{q_0} - i \frac{\hP}{p_0} \right).
\end{align}
where
\begin{align}\label{d300}
q_0 = \sqrt{\frac{\hbar}{m \omega}}, \qquad \text{and} \qquad p_0 = \frac{\hbar}{q_0} = \sqrt{m \omega \hbar}.
\end{align}
Inverting Eqs. \eqref{d380} we find
\begin{align}\label{d311bis}
\frac{\hQ}{q_0} = \frac{\ha + \had}{\sqrt{2} }, \qquad \frac{\hP}{p_0} =   \frac{\ha - \had}{\sqrt{2} \, i} .
\end{align}
It is not difficult to verify the validity of the following  relations:
\begin{align}
\bigl[ \ha, \had \bigr]   &  = 1, \label{HO10} \\[4pt]
\ha \ket{n} &  = \sqrt{n} \, \ket{n-1}, \label{HO20} \\[4pt]
\had \ket{n} &  = \sqrt{n+1} \, \ket{n+1}, \label{HO30} \\[4pt]
\mean{n'}{\hQ}{n} &  = \frac{q_0}{\sqrt{2}} \left( \sqrt{n} \, \delta_{n',n-1} + \sqrt{n+1} \, \delta_{n',n+1}  \right), \label{HO40} \\[4pt]
\mean{n'}{\hP}{n} &  = \frac{p_0}{\sqrt{2} \, i} \left( \sqrt{n} \, \delta_{n',n-1} -\sqrt{n+1} \, \delta_{n',n+1}  \right). \label{HO50}
\end{align}

The lowest energy eigenstate $\ket{n=0}$ can be written in coordinate and momentum representations as $\psi_Q(q) \deff  \brak{q_Q}{0}$ and $\psi_P(p) \deff  \brak{p_P}{0}$, respectively, where
\begin{align}
\psi_Q(q)  &  = \frac{1}{\pi^{1/4} \sqrt{q_0}} \exp \left( - \frac{q^2}{2 q_0^2}\right), \label{d280} \\[4pt]
\psi_P(p) &  = \frac{1}{\pi^{1/4} \sqrt{p_0}} \exp \left( - \frac{p^2}{2 p_0^2}\right), \label{d290}
\end{align}

From the spectral theorem \eqref{a200} and \eqref{a260}, it follows that given the vector $\ket{\psi} = \ket{0}$, there are two random variables, say $Q$ and $P$, associated with the operators $\hQ$ and $\hP$, respectively,  each following a Gaussian distribution, i.e.,
\begin{equation}\label{d302}
\begin{split}
Q \sim & \; f_{Q}^\psi(q) = \abs{\psi_Q(q)}^2 = \mathcal{N}_1(0, q_0^2/2) , \\[4pt]
P \sim &\; f_{P}^\psi(p) = \abs{\psi_P(p)}^2 = \mathcal{N}_1(0, p_0^2/2) .
\end{split}
\end{equation}
In the following, we will take
\begin{align}\label{d303}
\hX= \hQ/q_0, \qquad \text{and} \qquad \hY = \hP/p_0,
\end{align}
and $\ket{\psi} = \ket{0}$, so that
\begin{align}\label{d304}
\bigl[ \hX, \hY \bigr] = i \hI ,
\end{align}
and
\begin{equation}\label{m80}
\begin{split}
X \sim & \; f_{X}^\psi(x) =   \mathcal{N}_1(0, 1/2) , \\[4pt]
Y \sim & \; f_{Y}^\psi(y) =   \mathcal{N}_1(0, 1/2).
\end{split}
\end{equation}

\subsection{Sum of two Hermitian operators}\label{ContinuousExample2}

Here we calculate  $\ave{\hX + \hY}_\psi$ using both \eqref{c50} and  \eqref{c90} for  $\hX$ and $\hY$ given by \eqref{d303} and  $\ket{\psi} = \ket{0}$.

\subsubsection{Using the quasi-probability distribution $f(q,p)$}

\begin{figure}[ht!]
  \centering
  \includegraphics[scale=3,clip=false,width=1\columnwidth,trim = 0 0 0 0]{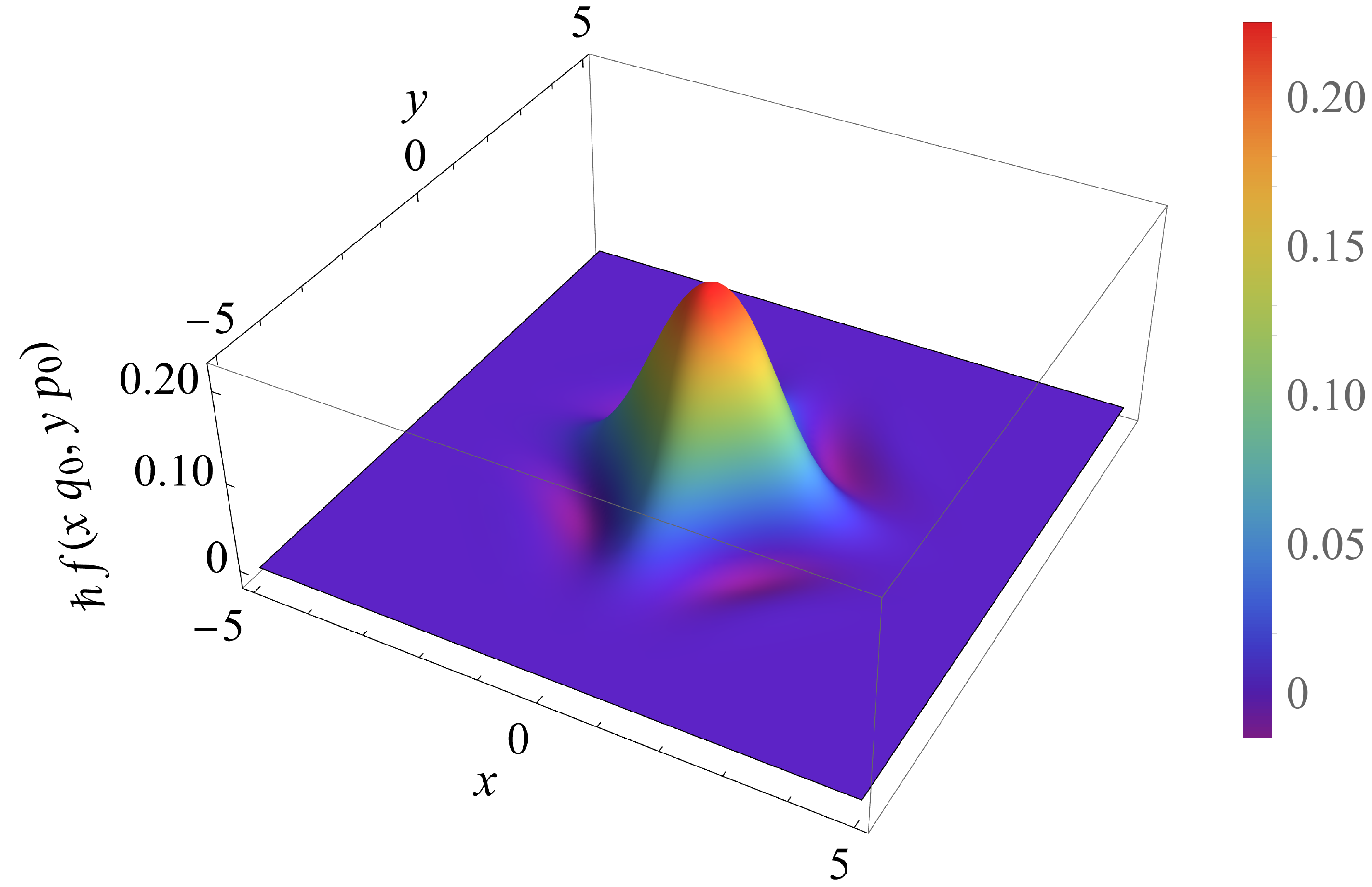}
  \caption{Plot of the quasi-probability distribution $\hbar \, f(x \, q_0, y \, p_0)$ given by Eq. \eqref{d310}. Viola-colored parts of the surface denote negative values of the function. }\label{fig2}
\end{figure}

From \eqref{d230}, \eqref{c60}   and \eqref{d275}  we have
\begin{align}\label{d309}
\hbar \, z(x \, q_0, y \, p_0) =  \frac{1}{\sqrt{2} \, \pi  } \exp  \left(- \frac{x^2 + y^2}{2}  + i \, x \, y \right),
\end{align}
so that
\begin{align}
\hbar \, f(x \, q_0, y \, p_0) = & \; \frac{1}{\sqrt{2} \, \pi  } \exp  \left(- \frac{x^2 + y^2}{2}  \right)   \cos \left( x y \right),  \label{d310} \\[4pt]
  \hbar \, g(x \, q_0, y \, p_0) = & \; \frac{1}{\sqrt{2} \, \pi} \exp \left(- \frac{x^2 + y^2}{2}  \right)   \sin \left( x y \right), \label{d310bis}
\end{align}
where $x$ and $y$ are dimensionless variables, and   \eqref{d280} and \eqref{d290} have been used.
Since $-1 \leq \cos \left( x \, y \right) \leq 1$, it is clear that $f(q,p)$ can take both positive and negative values, so it is not a PDF but a quasi-probability distribution. A plot of $f(q,p)$ is shown in Fig. \ref{fig2}.
%
%
%
%
%
%
In this case \eqref{c50}   gives correctly  $\mean{0}{\hX + \hY}{0}=0$,  because \eqref{d310} implies  $f(-q,-p) = f(q,p)$ and the integral of an odd function over a symmetric interval, namely the whole real axis $\mathbb{R}$, is equal to zero.

\subsubsection{Using the probability density function $f^\psi_W(w)$}

From \eqref{c90}  we determine  the characteristic function $\varphi_W^\psi(s) $ as
\begin{align}\label{m10}
\varphi_W^\psi(s) = \mean{0}{e^{i s (\hQ/q_0 + \hP/p_0)}}{0} .
\end{align}
To calculate this quantity we use the Campbell-Baker-Hausdorff theorem (see, for example, section 10.11.5 of Ref. \cite{MandelBook}), which states that
\begin{multline}\label{m20}
\exp \bigl[ \zeta ( \hA + \hB )\bigr] \\[4pt]
=   \exp ( \zeta  \hA )  \exp ( \zeta  \hB ) \exp\bigl( -\zeta^2 \bigl[\hA, \hB \bigr]/2 \bigr) \\[4pt]
=   \exp ( \zeta  \hB )  \exp ( \zeta  \hA ) \exp\bigl( \zeta^2 \bigl[\hA, \hB \bigr]/2 \bigr),
\end{multline}
for any $\zeta \in \mathbb{C}$, if
\begin{align}\label{m30}
\bigl[\hA, \bigl[\hA, \hB \bigr] \bigr] = 0 = \bigl[\hB, \bigl[\hA, \hB \bigr] \bigr].
\end{align}
We take $\zeta = i s$, $\hA = \hQ/q_0$ and $\hB = \hP/p_0$, to find
\begin{align}\label{m40}
\varphi_W^\psi(s) = e^{i s^2/2} \mean{0}{e^{i s \hQ/q_0} e^{i s \hP/p_0}}{0}.
\end{align}
We insert the identity $\hI=\int_\mathbb{R} \proj{q_Q}{q_Q} \, \di q$ just after the first exponential and the identity $\hI= \int_\mathbb{R} \proj{p_P}{p_P} \, \di p$  after the second exponential, and find
\begin{align}\label{m50}
\varphi_W^\psi(s) = & \;  e^{i s^2/2} \int_{\mathbb{R}^2} e^{i s(q/q_0 + p/p_0)} \nonumber \\[4pt]
& \times \brak{0}{q_Q} \brak{q_Q}{p_P} \brak{p_P}{0} \, \di q \, \di p \nonumber \\[4pt]
= & \;  e^{i s^2/2} \int_{\mathbb{R}^2} e^{i s(q/q_0 + p/p_0)} z(q,p) \, \di q \, \di p,
\end{align}
where \eqref{d230} has been used to write the last line.
Substituting \eqref{d309} into \eqref{m50} and performing the double integration, we obtain
\begin{align}\label{m60}
\varphi_W^\psi(s) = \exp \left(-  s^2 /2 \right).
\end{align}
Then, again from \eqref{c90} we find
\begin{align}\label{m70}
f_W^\psi(w)  = & \; \frac{1}{2 \pi} \int_{\mathbb{R}} e^{- i s w}  \varphi_W^\psi(s) \, \di s  \nonumber \\[4pt]
 = & \;  \frac{1}{\sqrt{2 \pi}} \exp \left(-  w^2 /2 \right) \nonumber \\[4pt]
 = & \; \mathcal{N}_1(0,1).
\end{align}
As the expected value of a random variable following the normal distribution $\mathcal{N}_1(0,1)$ is zero, we find again $\mean{0}{\hX + \hY}{0}=0$.

\subsubsection{Discussion}

It is instructive to compare the results above with $f_W(w)$ and $\varphi_W(s)$ given by \eqref{e100} and \eqref{delta40}. To this end, we must first match \eqref{e20bis} with \eqref{m80} by choosing
 $\sigma_1^2 = 1/2 = \sigma_2^2$, and $\mu_1 = 0 =\mu_2$. Next, if we also choose $\rho = 0$ (i.e., absence of correlation) into \eqref{e100} and \eqref{delta40}, we find
\begin{equation}\label{m900}
\begin{split}
\varphi_W^\psi(s) =  & \; \varphi_W(s) , \\[4pt]
f_W^\psi(w) = & \; f_W(w).
\end{split}
\end{equation}
Given this pleasant result, one may be tempted  thinking that there could be a one-to-one correspondence between quantum and classical PDFs. However, the next example shows that this is not the case.

\subsection{Product of two Hermitian operators}\label{ContinuousExample1}

Our task here is  calculating $\ave{\hU}_\psi$ and  $\ave{\hV}_\psi$ using both (\ref{d220}-\ref{d225})  and (\ref{d220bis}-\ref{d225bis}), again for  $\hX$ and $\hY$ given by \eqref{d303} and  $\ket{\psi} = \ket{0}$.

\subsubsection{Using $f(q,p)$ and $g(q,p)$}

From \eqref{d200} and \eqref{d303} we have
\begin{align}
\hU  = & \; \frac{\hQ \hP + \hP \hQ}{2  \hbar} , \label{m100} \\[4pt]
\hV  = & \; \frac{\hQ \hP - \hP \hQ}{2 i  \hbar} = \frac{\hI}{2} , \label{m110}
\end{align}
because by definition $q_0 p_0 =\hbar$, and \eqref{d268} has been used. We remark that $\hU$ is proportional to the sum of two non-Hermitian operators. However, it is easy to check that $\hQ \hP$ is a so-called \emph{normal operator}, that is an operator that commutes with its Hermitian conjugate: $\bigl[ \hQ \hP , (\hQ \hP)^\dagger \bigr] = \bigl[ \hQ \hP , \hP \hQ  \bigr] = 0$.
Using $f(q,p)$ from \eqref{d310} and $g(q,p)$ from \eqref{d310bis}, after a straightforward integration we obtain
\begin{align}
\ave{\hU}_\psi = & \;  0, \label{d320} \\[4pt]
\ave{\hV}_\psi = & \; \frac{1}{2}. \label{d322}
\end{align}
From $\ave{\hP \hQ}_\psi = \ave{\hQ \hP}_\psi^*$ and \eqref{d322}, it follows that $\ave{\bigl[ \hQ , \hP \bigr]}_\psi = i \hbar$, as it should be.

\subsubsection{Using the probability density functions $f^\psi_U(u)$ and $f^\psi_V(v)$}

From \eqref{d312} and \eqref{d304} it follows that the  calculation of $f^\psi_V(v)$ is very easy because
\begin{align}\label{d370}
f^\psi_V(v) = & \; \mean{0}{\delta \bigl( v -  \hI/2 \bigr)}{0} \nonumber \\[4pt]
 = & \; \delta \left( v - {1}/{2}  \right).
\end{align}
Substituting this result into \eqref{d225bis} we obtain immediately $\ave{\hV}_\psi = {1}/{2}$.

Calculation of $f^\psi_U(u)$ is more involved because it requires the use of the annihilation and creation operators $\ha$ and $\had$ defined by \eqref{d380}.
From \eqref{m100} and \eqref{d311bis}, we have
\begin{align}\label{d400}
\hU = \frac{\hQ \hP + \hP \hQ}{2 \, \hbar} = \frac{1}{2 i}\left( \ha^2 - \hadb \right).
\end{align}
Next we substitute \eqref{d400} into \eqref{d311} and use \eqref{a370} to obtain
\begin{align}\label{d410}
f^\psi_U(u) = & \; \mean{0}{\frac{1}{2 \pi} \int_{\mathbb{R}}  e^{-i s \left[ u - \frac{1}{2 i}\left( \ha^2 - \hadb \right) \right]} \,  \di s}{0} \nonumber \\[4pt]
= & \; \frac{1}{2 \pi} \int_{\mathbb{R}}  e^{-i s u} \mean{0}{
 e^{\frac {s }{2 } \left( \ha^2 - \hadb \right) } }{0}\,  \di s.
\end{align}
The operator
\begin{align}\label{d420}
\hat{S}(\zeta) = \exp \left( \frac{\zeta^*}{2} \ha^2 - \frac{\zeta}{2} \hadb \right),
\end{align}
where $\zeta = r \exp(i \varphi) \in \mathbb{C}, ~ r \geq 0, ~ \varphi \in \mathbb{R}$,
is known as the \emph{squeezing operator} in quantum optics, and the state
\begin{align}\label{d430}
\ket{\zeta} = & \; \hat{S}(\zeta) \ket{0} \nonumber \\[4pt]
= & \; \frac{1}{\sqrt{\cosh r}} \, \sum_{n=0}^\infty \frac{\sqrt{(2n)!}}{n!} \left(-\frac{e^{i \varphi}}{2} \tanh r\right)^n \ket{2 n},
\end{align}
 is termed the \emph{squeezed vacuum} (see, for example, section 3.7 of Ref. \cite{Barnett}). Here $\ket{2n}$ is the eigenvector of the harmonic oscillator Hamiltonian \eqref{d262} associated with the eigenvalue $E_{2n}$. Using \eqref{d430} we can straightforwardly calculate from \eqref{d410},
\begin{align}\label{d440}
f^\psi_U(u) = & \; \frac{1}{2 \pi} \int_{\mathbb{R}}  e^{-i s u} \mean{0}{
\hat{S}(s  ) }{0}\,  \di s \nonumber \\[4pt]
= & \; \frac{1}{2 \pi} \int_{\mathbb{R}}  e^{-i s u} \frac{1}{\sqrt{\cosh s}}   \,  \di s  ,
\end{align}
where we have used the orthogonality condition $\brak{0}{2n} = \delta_{0n}$. It is not difficult to show that $f^\psi_U(u)$ is correctly normalized. To this end, we must evaluate
\begin{align}\label{d442}
\int_\mathbb{R} f^\psi_U(u) \, \di u = & \; \frac{1}{2 \pi} \int_{\mathbb{R}^2}  e^{-i s u} \frac{1}{\sqrt{\cosh s}}  \,  \di s \, \di u.
\end{align}
We carry out first the $u$ integration using \eqref{a80h}, thus obtaining a delta distribution $\delta(s)$. This yields
\begin{align}\label{d443}
\int_\mathbb{R} f^\psi_U(u) \, \di u = & \;  \int_{\mathbb{R}} \frac{1}{\sqrt{\cosh s}}   \, \delta(s) \, \di s  \nonumber \\[4pt]
= & \; 1.
\end{align}
Moreover,  $\cosh(-s) = \cosh(s)$ implies both $f^\psi_U(u) \in \mathbb{R}$,  and $f^\psi_U(-u) = f^\psi_U(u)$.

To find $\ave{\hU}_\psi$ actually we do not need to calculate explicitly the integral in  \eqref{d440} because from the symmetry $f^\psi_U(-u) = f^\psi_U(u)$ it follows automatically that all the odd momenta of $f^\psi_U(u)$ are zero, so that $\ave{\hU}_\psi =  0$. However, it is more instructive to note that
from the definition of quantum characteristic function via \eqref{a372}, we have
\begin{align}\label{d460}
\varphi_U^\psi(s) \deff  \frac{1}{\sqrt{\cosh s}} .
\end{align}
Finally, from \eqref{a84} we find
\begin{align}\label{d470}
\ave{\hU}_\psi = & \; \left. \frac{1}{i} \frac{\di \varphi_U^\psi(s)}{\di s} \right|_{s=0} \nonumber \\[4pt]
= & \; \left. \frac{i}{2}  \frac{\tanh s}{\sqrt{\cosh s}}  \, \right|_{s=0}  =0 . 
%
%
\end{align}

\begin{figure}[h!]
  \centering
  \includegraphics[scale=3,clip=false,width=1\columnwidth,trim = 0 0 0 0]{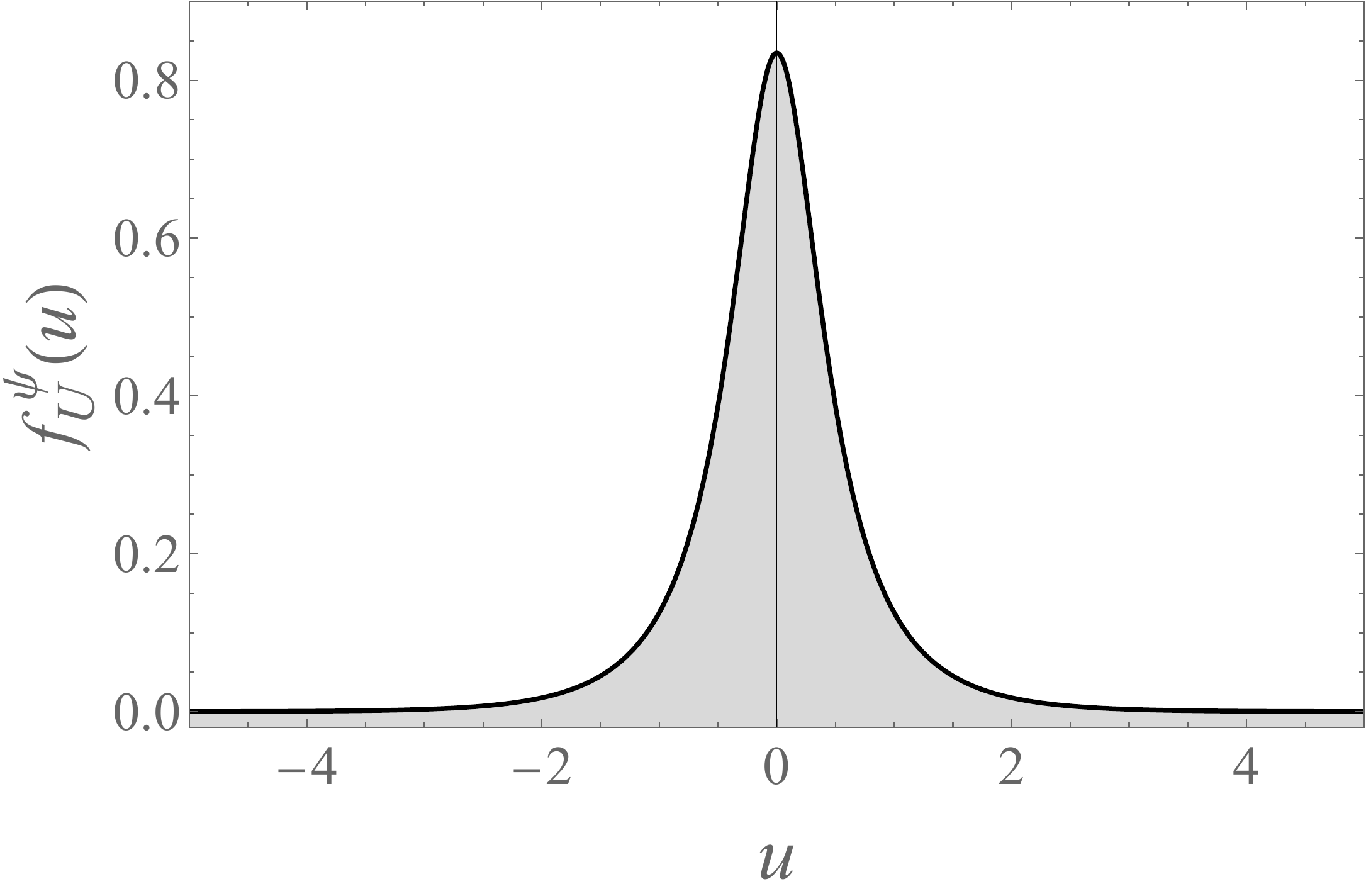}
  \caption{Plot of the probability density function $f_U^\psi(u)$ given by Eq. \eqref{d480}.}\label{fig3}
\end{figure}

For completeness, we remark that the integral in Eq. \eqref{d440} can be actually calculated. Mathematica gives
\begin{align}\label{d480}
f_U^\psi(u) = & \; \frac{e^{i \pi /4}}{2^{3/2}\pi} \left[ \exp \left(\frac{\pi  u}{2} \right) B_{-1}\left(\frac{1}{4} + i \frac{u}{2} ,\frac{1}{2}\right) \right. \nonumber \\[4pt]
& \left. + \exp \left(-\frac{\pi  u}{2} \right) B_{-1}\left(\frac{1}{4}- i \frac{ u}{2},\frac{1}{2}\right) \right],
\end{align}
where $B_z(a,b)$ denotes the \emph{incomplete beta function} \cite{IncompleteBetaFunction}. A plot of this function is shown in Fig. \ref{fig3}
%
%
%
%
%
%

\subsubsection{Discussion}

In this case, comparison with the results of Sec. \ref{Two_examples} is a somewhat more subtle affair. Let us try to give first a brief summary of the situation. Given  the operators $\hX$ and $\hY$ defined by \eqref{d303}, there are two real-valued random variables $X \sim \mathcal{N}_1(0,1/2)$ and $Y \sim \mathcal{N}_1(0,1/2)$ defined by \eqref{m80}, which \emph{statistically} reproduce the results of \emph{separate}, as opposed to joint, measurements of the observables represented by $\hX$ and $\hY$, made on a quantum harmonic oscillator prepared in the ground state $\ket{0}$. In Sec. \ref{Two_examples} we have considered the two-dimensional random vector $\bX = (X_1,X_2)$, characterized by the joint normal distribution $\bX \sim \mathcal{N}_2(\bmu,V)$. Each of two components $X_1$ and $X_2$ of $\bX$  follows a marginal normal distribution: $X_n \sim \mathcal{N}_1(\mu_n,\sigma_n^2),~(n=1,2)$. Then we have shown that the \emph{new} random variable $Y = X_1 X_2$ is characterized by the PDF $f_Y(y)$ defined by \eqref{e60} and drawn in Fig. \ref{fig1}.

In quantum mechanics the operator $\hX \hY$ is not Hermitian, therefore it does not describe any physical observable of the harmonic oscillator and the spectral theorem does not apply. Thus, while from \eqref{e30} we have $\ave{X_1 X_2} = \rho \, \sigma_1 \sigma_2 \in \mathbb{R}$, now from \eqref{f30bis} and (\ref{HO40}-\ref{HO50}) we have
\begin{align}\label{q10}
\mean{0}{\hX \hY}{0} = & \; \mean{0}{\hX \hI \hY}{0} \nonumber \\[4pt]
 = & \; \sum_{n=0}^\infty \mean{0}{\hX}{n}\mean{n}{\hY}{0} \nonumber \\[4pt]
 = & \; \frac{1}{q_0 p_0} \sum_{n=0}^\infty \mean{0}{\hQ}{n} \bigl( \mean{0}{\hP}{n} \bigr)^* \nonumber \\[4pt]
 = & \; \frac{1}{q_0 p_0} \sum_{n=0}^\infty \frac{q_0}{\sqrt{2}}  \sqrt{n} \, \delta_{0,n-1} \left( \frac{p_0}{\sqrt{2} \, i}  \sqrt{n} \, \delta_{0,n-1} \right)^* \nonumber \\[4pt]
 = & \; \frac{i}{2} \in \mathbb{C}.
\end{align}

\begin{figure}[ht!]
  \centering
  \includegraphics[scale=3,clip=false,width=1\columnwidth,trim = 0 0 0 0]{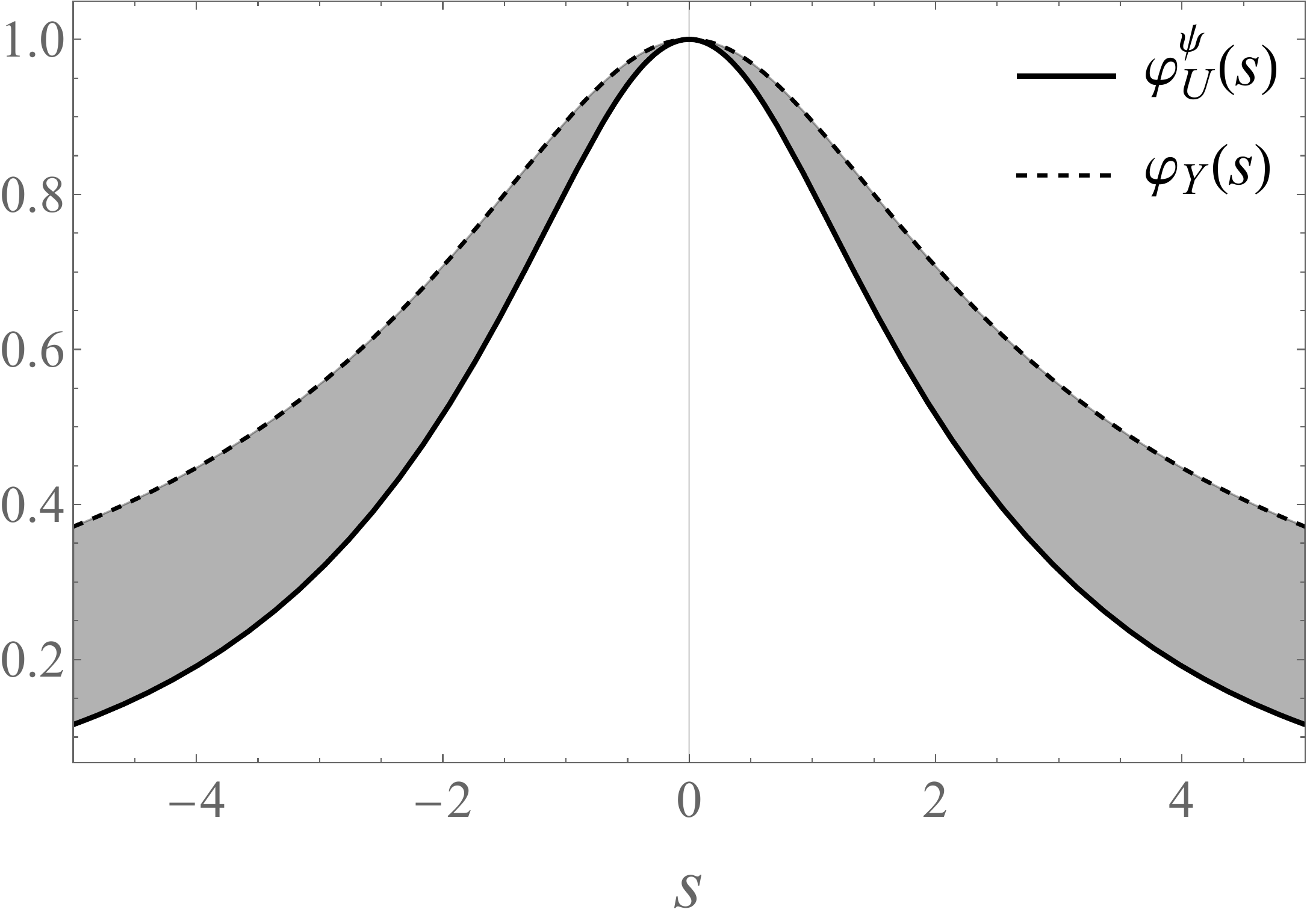}
  \caption{Comparison between the two  characteristic functions $\varphi_U^\psi(s) = (\cosh s)^{-1/2}$ and $\varphi_Y(s) = (1+s^2/4)^{-1/2}$. The latter has been calculated from \eqref{delta50} with $\sigma_1^2 = 1/2 = \sigma_2^2$, $\mu_1 = 0 = \mu_2$ and $\rho =0$. The gray area is meant to highlight the difference between the two curves. }\label{fig4}
\end{figure}

Therefore, we have introduced the two Hermitian operators $\hU$ and $\hV$ defined by $\hX \hY = \hU + i \, \hV $ and (\ref{m100}-\ref{m110}). Next, we have applied the spectral theorem to $\hU$ and $\hV$ separately and we have found two real-valued random variables $U$ and $V$ such that (see \eqref{d370} and \eqref{d440}),
\begin{align}
U \sim  & \; f_U^\psi(u) =  \frac{1}{2 \pi} \int_{\mathbb{R}}  e^{-i s u} \frac{1}{\sqrt{\cosh s}}   \,  \di s  , \label{q20} \\[4pt]
V \sim  & \;  f_V^\psi(v) =  \delta(v - 1/2) .\label{q30}
\end{align}
First, we note that $V$ is actually a discrete random variable. In fact, from \eqref{q30} and the definition \eqref{a52} of probability mass function it follows that
\begin{align}\label{q40}
p_V(v) = \prob{V = v} = \left\{
                          \begin{array}{ll}
                            1, & v = 1/2, \\[4pt]
                            0, & v \neq 1/2.
                          \end{array}
                        \right.
\end{align}
So, there is no connection whatsoever between $Y = X_1 X_2$ and $V$. Actually, the latter is not a random quantity at all, since its only value $V = 1/2$ occurs with certainty.
Second, the random variable $U$ is characterized by the PDF $f_U(u)$ depicted in Fig. \ref{fig2}, which bears little resemblance with $f_Y(y)$  plotted in  Fig. \ref{fig2}, apart from the same symmetry when $\rho = 0$. %
It may be perhaps more useful to visually compare the characteristic functions $\varphi_Y(s)$ defined by \eqref{delta50}, and $\varphi_U^\psi(s)$ defined by \eqref{d460}, provided that the marginal PDFs \eqref{e20bis} and \eqref{m80} coincide by choosing $\sigma_1^2 = 1/2 = \sigma_2^2$ and $\mu_1 = 0 = \mu_2$. Furthermore, since $\varphi_Y(s) \in \mathbb{C}$ for $\rho \neq 0$, we must also choose $\rho =0$. Both functions are shown in Fig. \ref{fig4} for comparison.
%
%
%
%
%
%
So, in the end the connection  between $Y = X_1 X_2$ and $U$ is not so clear.

The results presented in this section show that little can be said with certainty about the correspondence between Hermitian quantum operators and real-valued random variables provided by the spectral theorem, when the  operators are functions of noncommuting operators describing incompatible observables, because the latter cannot have definite values simultaneously.

\section{Concluding remarks}\label{conclusions}

In this paper we provided some examples of the way in which the rules of classical and quantum probability agree or differ. The main goal was to provide students with conceptual and calculational tools that would permit them to gain a deeper understanding of quantum mechanics.
We have shown that there is a complete agreement between classical and quantum probability distributions in the case of compatible observables. Conversely, we have found either agreement or disagreement in the case of  incompatible quantum observables, depending on the specific observables and the quantum state of the system.

In conclusion,  considering quantum mechanics from the point of view of probability distributions (see also Ref. \cite{OperationalQP} for a similar approach), gives us an additional and valuable view of the comparison between classical and quantum physics. In particular, this approach shows  how the existence of incompatible observables underlies the profound differences between the two mechanics.

\begin{acknowledgments}

I acknowledge support from Deutsche Forschungsgemeinschaft Project No. 429529648-
TRR 306 QuCoLiMa (``Quantum Cooperativity of Light and Matter''). I thank Peter Morgan for bringing to my attention the work \cite{Morgan_2022}, where some of the techniques presented here are used. 

\end{acknowledgments}


\begin{thebibliography}{36}
\expandafter\ifx\csname natexlab\endcsname\relax\def\natexlab#1{#1}\fi
\expandafter\ifx\csname bibnamefont\endcsname\relax
  \def\bibnamefont#1{#1}\fi
\expandafter\ifx\csname bibfnamefont\endcsname\relax
  \def\bibfnamefont#1{#1}\fi
\expandafter\ifx\csname citenamefont\endcsname\relax
  \def\citenamefont#1{#1}\fi
\expandafter\ifx\csname url\endcsname\relax
  \def\url#1{\texttt{#1}}\fi
\expandafter\ifx\csname urlprefix\endcsname\relax\def\urlprefix{URL }\fi
\providecommand{\bibinfo}[2]{#2}
\providecommand{\eprint}[2][]{\url{#2}}

\bibitem[{\citenamefont{Dragoman and Dragoman}(2004)}]{Dragoman}
\bibinfo{author}{\bibfnamefont{D.}~\bibnamefont{Dragoman}} \bibnamefont{and}
  \bibinfo{author}{\bibfnamefont{M.}~\bibnamefont{Dragoman}},
  \emph{\bibinfo{title}{{Quantum-Classical Analogies}}}, The Frontiers
  Collection (\bibinfo{publisher}{Springer Berlin, Heidelberg},
  \bibinfo{year}{2004}), \bibinfo{edition}{1st} ed., ISBN
  \bibinfo{isbn}{978-3-662-09647-5}.

\bibitem[{\citenamefont{von Neumann}(2018)}]{vonNeumann2018}
\bibinfo{author}{\bibfnamefont{J.}~\bibnamefont{von Neumann}},
  \emph{\bibinfo{title}{Mathematical Foundations of Quantum Mechanics}}
  (\bibinfo{publisher}{Princeton University Press},
  \bibinfo{address}{Princeton}, \bibinfo{year}{2018}), ISBN
  \bibinfo{isbn}{9781400889921},
  \urlprefix\url{https://doi.org/10.1515/9781400889921}.

\bibitem[{\citenamefont{{Frigyes Riesz} and {B\'{e}la
  Sz.-Nagy}}(1990)}]{Riesz_Nagy}
\bibinfo{author}{\bibnamefont{{Frigyes Riesz}}} \bibnamefont{and}
  \bibinfo{author}{\bibnamefont{{B\'{e}la Sz.-Nagy}}},
  \emph{\bibinfo{title}{{Functional Analysis}}} (\bibinfo{publisher}{Dover
  Publications, Inc.}, \bibinfo{address}{New York}, \bibinfo{year}{1990}), ISBN
  \bibinfo{isbn}{0-486-66289-6}.

\bibitem[{\citenamefont{Dirac}(1981)}]{Dirac}
\bibinfo{author}{\bibfnamefont{P.~A.~M.} \bibnamefont{Dirac}},
  \emph{\bibinfo{title}{The Principles of Quantum Mechanics}}
  (\bibinfo{publisher}{Clarendon Press}, \bibinfo{address}{Oxford},
  \bibinfo{year}{1981}), \bibinfo{edition}{4th} ed.

\bibitem[{\citenamefont{Sakurai}(1994)}]{Sakurai}
\bibinfo{author}{\bibfnamefont{J.~J.} \bibnamefont{Sakurai}},
  \emph{\bibinfo{title}{{Modern quantum mechanics; rev. ed.}}}
  (\bibinfo{publisher}{Addison-Wesley}, \bibinfo{address}{Reading, MA},
  \bibinfo{year}{1994}).

\bibitem[{\citenamefont{Kronz and Lupher}(2021)}]{sep-qt-nvd}
\bibinfo{author}{\bibfnamefont{F.}~\bibnamefont{Kronz}} \bibnamefont{and}
  \bibinfo{author}{\bibfnamefont{T.}~\bibnamefont{Lupher}}, in
  \emph{\bibinfo{booktitle}{The {Stanford} Encyclopedia of Philosophy}}, edited
  by \bibinfo{editor}{\bibfnamefont{E.~N.} \bibnamefont{Zalta}}
  (\bibinfo{publisher}{Metaphysics Research Lab, Stanford University},
  \bibinfo{year}{2021}), \bibinfo{edition}{{W}inter 2021} ed.

\bibitem[{\citenamefont{Inc.}()}]{Mathematica}
\bibinfo{author}{\bibfnamefont{W.~R.} \bibnamefont{Inc.}},
  \emph{\bibinfo{title}{Mathematica, {V}ersion 13.1}},
  \bibinfo{note}{{Champaign, IL, 2022}},
  \urlprefix\url{https://www.wolfram.com/mathematica}.

\bibitem[{\citenamefont{Horn and Johnson}(1985)}]{horn_johnson_1985}
\bibinfo{author}{\bibfnamefont{R.~A.} \bibnamefont{Horn}} \bibnamefont{and}
  \bibinfo{author}{\bibfnamefont{C.~R.} \bibnamefont{Johnson}},
  \emph{\bibinfo{title}{Matrix Analysis}} (\bibinfo{publisher}{Cambridge
  University Press}, \bibinfo{address}{Cambridge}, \bibinfo{year}{1985}), ISBN
  \bibinfo{isbn}{978-0-521-38632-2}.

\bibitem[{\citenamefont{Lighthill}(1958)}]{lighthill_1958}
\bibinfo{author}{\bibfnamefont{M.~J.} \bibnamefont{Lighthill}},
  \emph{\bibinfo{title}{An Introduction to Fourier Analysis and Generalised
  Functions}}, Cambridge Monographs on Mechanics (\bibinfo{publisher}{Cambridge
  University Press}, \bibinfo{address}{Cambridge}, \bibinfo{year}{1958}).

\bibitem[{\citenamefont{Weisstein}({\natexlab{a}})}]{Hfunction}
\bibinfo{author}{\bibfnamefont{E.~W.} \bibnamefont{Weisstein}},
  \emph{\bibinfo{title}{``{Heaviside Step Function.}''}},
  \bibinfo{howpublished}{From \emph{MathWorld}--A Wolfram Web Resource},
  \urlprefix\url{https://mathworld.wolfram.com/HeavisideStepFunction.html}.

\bibitem[{\citenamefont{Weisstein}({\natexlab{b}})}]{DeltaFunction}
\bibinfo{author}{\bibfnamefont{E.~W.} \bibnamefont{Weisstein}},
  \emph{\bibinfo{title}{``{Delta Function.}''}}, \bibinfo{howpublished}{From
  \emph{MathWorld}--A Wolfram Web Resource},
  \urlprefix\url{https://mathworld.wolfram.com/DeltaFunction.html}.

\bibitem[{Sti()}]{Stieltjes}
\bibinfo{note}{For our purposes it is not important to explain what a Stieltjes
  integral is because we will never use it in practical calculations. The
  interested reader can find a non-mathematical definition of the Stieltjes
  integral in section 1.3 of Ref. \cite{Ballentine}, and a rigorous
  mathematical one in Ref. \cite{Riesz_Nagy}.}

\bibitem[{\citenamefont{Feller}(1971)}]{FellerII}
\bibinfo{author}{\bibfnamefont{W.}~\bibnamefont{Feller}},
  \emph{\bibinfo{title}{{An Introduction to Probability Theory and Its
  Applications}}}, vol.~\bibinfo{volume}{II} (\bibinfo{publisher}{John Wiley \&
  Sons, Inc.}, \bibinfo{year}{1971}), \bibinfo{edition}{2nd} ed., ISBN
  \bibinfo{isbn}{0-471-25709-5}.

\bibitem[{\citenamefont{{Papoulis, A.} and {Pillai, S.
  U.}}(2002)}]{papoulis2002}
\bibinfo{author}{\bibnamefont{{Papoulis, A.}}} \bibnamefont{and}
  \bibinfo{author}{\bibnamefont{{Pillai, S. U.}}},
  \emph{\bibinfo{title}{{Probability, Random Variables, and Stochastic
  Processes}}} (\bibinfo{publisher}{McGraw-Hill}, \bibinfo{year}{2002}),
  \bibinfo{edition}{4th} ed., ISBN \bibinfo{isbn}{0-07-366011-6}.

\bibitem[{\citenamefont{Taboga}(2017)}]{taboga2017}
\bibinfo{author}{\bibfnamefont{M.}~\bibnamefont{Taboga}},
  \emph{\bibinfo{title}{{Lectures on Probability Theory and Mathematical
  Statistics}}} (\bibinfo{publisher}{CreateSpace Independent Publishing
  Platform}, \bibinfo{year}{2017}), \bibinfo{edition}{3rd} ed., ISBN
  \bibinfo{isbn}{9781981369195}.

\bibitem[{\citenamefont{Gillespie}(1983)}]{Gillespie1983}
\bibinfo{author}{\bibfnamefont{D.~T.} \bibnamefont{Gillespie}},
  \bibinfo{journal}{American Journal of Physics} \textbf{\bibinfo{volume}{51}},
  \bibinfo{pages}{520} (\bibinfo{year}{1983}), ISSN \bibinfo{issn}{0002-9505},
  \urlprefix\url{https://doi.org/10.1119/1.13221}.

\bibitem[{\citenamefont{Ramshaw}(1985)}]{Ramshaw1985}
\bibinfo{author}{\bibfnamefont{J.~D.} \bibnamefont{Ramshaw}},
  \bibinfo{journal}{American Journal of Physics} \textbf{\bibinfo{volume}{53}},
  \bibinfo{pages}{178} (\bibinfo{year}{1985}), ISSN \bibinfo{issn}{0002-9505},
  \urlprefix\url{https://doi.org/10.1119/1.14109}.

\bibitem[{\citenamefont{Paveri-Fontana}(1991)}]{Paveri-Fontana1991}
\bibinfo{author}{\bibfnamefont{S.~L.} \bibnamefont{Paveri-Fontana}},
  \bibinfo{journal}{American Journal of Physics} \textbf{\bibinfo{volume}{59}},
  \bibinfo{pages}{854} (\bibinfo{year}{1991}),
  \urlprefix\url{https://doi.org/10.1119/1.16739}.

\bibitem[{\citenamefont{Weisstein}({\natexlab{c}})}]{ModifiedBesselSec}
\bibinfo{author}{\bibfnamefont{E.~W.} \bibnamefont{Weisstein}},
  \emph{\bibinfo{title}{``{Modified Bessel Function of the Second Kind.}''}},
  \bibinfo{howpublished}{From \emph{MathWorld}--A Wolfram Web Resource},
  \urlprefix\url{https://mathworld.wolfram.com/ModifiedBesselFunctionoftheSecondKind.html}.

\bibitem[{Not({\natexlab{a}})}]{NoteHermitian}
\bibinfo{note}{Following Sakurai's book, throughout this paper we will write
  ``Hermitian operators'' to denote both Hermitian matrices and self-adjoint
  linear operators.}

\bibitem[{\citenamefont{Jordan}(1997)}]{jordan1997}
\bibinfo{author}{\bibfnamefont{T.~F.} \bibnamefont{Jordan}},
  \emph{\bibinfo{title}{{Linear Operators for Quantum Mechanics}}}
  (\bibinfo{publisher}{Dover Publications, Inc.}, \bibinfo{address}{Mineola,
  New York}, \bibinfo{year}{1997}), ISBN \bibinfo{isbn}{978-0-486-45329-3}.

\bibitem[{\citenamefont{{Leslie E. Ballentine}}(1998)}]{Ballentine}
\bibinfo{author}{\bibnamefont{{Leslie E. Ballentine}}},
  \emph{\bibinfo{title}{Quantum Mechanics: A Modern Development}}
  (\bibinfo{publisher}{World Scientific Publishing},
  \bibinfo{address}{Singapore}, \bibinfo{year}{1998}), ISBN
  \bibinfo{isbn}{981-02-4105-4}.

\bibitem[{Not({\natexlab{b}})}]{NoteSakurai}
\bibinfo{note}{See section 1.6 of \cite{Sakurai} for a quick review. Following
  Sakurai's pedagogical spirit, we will talk about eigenvectors and eigenvalues
  of an operator with a purely continuous spectrum even when they, rigorously
  speaking, do not exist. For a gentle introduction to this topic, the
  interested reader can consult de la Madrid's didactic article
  \cite{de_la_Madrid2005}.}

\bibitem[{\citenamefont{{Alberto Galindo} and {Pedro
  Pascual}}(1990)}]{GalindoI}
\bibinfo{author}{\bibnamefont{{Alberto Galindo}}} \bibnamefont{and}
  \bibinfo{author}{\bibnamefont{{Pedro Pascual}}},
  \emph{\bibinfo{title}{Quantum Mechanics I}}, {Texts and Monographs in Physics
  (TMP)} (\bibinfo{publisher}{Springer-Verlag}, \bibinfo{address}{Berlin,
  Heidelberg}, \bibinfo{year}{1990}).

\bibitem[{Col()}]{Coleman2019b}
\bibinfo{note}{See problem \textbf{4.3} in Ref. \cite{Coleman2019}.}

\bibitem[{\citenamefont{Xu et~al.}(2021)\citenamefont{Xu, Ding, Peng, and
  Fang}}]{Xu2021}
\bibinfo{author}{\bibfnamefont{Y.-J.} \bibnamefont{Xu}},
  \bibinfo{author}{\bibfnamefont{X.-C.} \bibnamefont{Ding}},
  \bibinfo{author}{\bibfnamefont{J.-Z.} \bibnamefont{Peng}}, \bibnamefont{and}
  \bibinfo{author}{\bibfnamefont{S.-D.} \bibnamefont{Fang}},
  \bibinfo{journal}{American Journal of Physics} \textbf{\bibinfo{volume}{89}},
  \bibinfo{pages}{535} (\bibinfo{year}{2021}), ISSN \bibinfo{issn}{0002-9505},
  \urlprefix\url{https://doi.org/10.1119/10.0003900}.

\bibitem[{\citenamefont{Case}(2008)}]{Case2008}
\bibinfo{author}{\bibfnamefont{W.~B.} \bibnamefont{Case}},
  \bibinfo{journal}{American Journal of Physics} \textbf{\bibinfo{volume}{76}},
  \bibinfo{pages}{937} (\bibinfo{year}{2008}), ISSN \bibinfo{issn}{0002-9505},
  \urlprefix\url{https://doi.org/10.1119/1.2957889}.

\bibitem[{\citenamefont{L{\'e}vy-Leblond}(1986)}]{Levy-Leblond1986}
\bibinfo{author}{\bibfnamefont{J.}~\bibnamefont{L{\'e}vy-Leblond}},
  \bibinfo{journal}{American Journal of Physics} \textbf{\bibinfo{volume}{54}},
  \bibinfo{pages}{135} (\bibinfo{year}{1986}), ISSN \bibinfo{issn}{0002-9505},
  \urlprefix\url{https://doi.org/10.1119/1.14708}.

\bibitem[{\citenamefont{Campos}(1998)}]{Campos1998}
\bibinfo{author}{\bibfnamefont{R.~A.} \bibnamefont{Campos}},
  \bibinfo{journal}{American Journal of Physics} \textbf{\bibinfo{volume}{66}},
  \bibinfo{pages}{712} (\bibinfo{year}{1998}), ISSN \bibinfo{issn}{0002-9505},
  \urlprefix\url{https://doi.org/10.1119/1.18937}.

\bibitem[{\citenamefont{{L. Mandel} and {E. Wolf}}(1995)}]{MandelBook}
\bibinfo{author}{\bibnamefont{{L. Mandel}}} \bibnamefont{and}
  \bibinfo{author}{\bibnamefont{{E. Wolf}}}, \emph{\bibinfo{title}{Optical
  Coherence and Quantum Optics}} (\bibinfo{publisher}{Cambridge University
  Press}, \bibinfo{address}{New York}, \bibinfo{year}{1995}).

\bibitem[{\citenamefont{{Stephen M. Barnett} and {Paul M.
  Radmore}}(2002)}]{Barnett}
\bibinfo{author}{\bibnamefont{{Stephen M. Barnett}}} \bibnamefont{and}
  \bibinfo{author}{\bibnamefont{{Paul M. Radmore}}},
  \emph{\bibinfo{title}{{Methods in theoretical Quantum Optics}}}, Oxford
  Series in Optical and Imaging Science: 15 (\bibinfo{publisher}{Oxford
  University Press}, \bibinfo{address}{Oxford, UK}, \bibinfo{year}{2002}), ISBN
  \bibinfo{isbn}{0 19 856361 2}.

\bibitem[{\citenamefont{Weisstein}({\natexlab{d}})}]{IncompleteBetaFunction}
\bibinfo{author}{\bibfnamefont{E.~W.} \bibnamefont{Weisstein}},
  \emph{\bibinfo{title}{``{Incomplete Beta Function.}''}},
  \bibinfo{howpublished}{From \emph{MathWorld}--A Wolfram Web Resource},
  \urlprefix\url{https://mathworld.wolfram.com/IncompleteBetaFunction.html}.

\bibitem[{\citenamefont{{Paul Busch} et~al.}(1995)\citenamefont{{Paul Busch},
  {Marian Grabowski}, and {Pekka J. Lahti}}}]{OperationalQP}
\bibinfo{author}{\bibnamefont{{Paul Busch}}},
  \bibinfo{author}{\bibnamefont{{Marian Grabowski}}}, \bibnamefont{and}
  \bibinfo{author}{\bibnamefont{{Pekka J. Lahti}}},
  \emph{\bibinfo{title}{Operational Quantum Physics}}, Lecture Notes in Physics
  Monographs (\bibinfo{publisher}{Springer Berlin, Heidelberg},
  \bibinfo{year}{1995}).



\bibitem[{\citenamefont{de~la Madrid}(2005)}]{de_la_Madrid2005}
\bibinfo{author}{\bibfnamefont{R.}~\bibnamefont{de~la Madrid}},
  \bibinfo{journal}{European Journal of Physics} \textbf{\bibinfo{volume}{26}},
  \bibinfo{pages}{287} (\bibinfo{year}{2005}),
  \urlprefix\url{https://dx.doi.org/10.1088/0143-0807/26/2/008}.

\bibitem[{\citenamefont{Coleman}(2019)}]{Coleman2019}
\bibinfo{author}{\bibfnamefont{S.}~\bibnamefont{Coleman}},
  \emph{\bibinfo{title}{Quantum Field Theory}} (\bibinfo{publisher}{World
  Scientific Publishing}, \bibinfo{address}{Singapore}, \bibinfo{year}{2019}),
  ISBN \bibinfo{isbn}{978-981-4635-50-9}.
  
  \bibitem[{\citenamefont{Morgan}(2022)}]{Morgan_2022}
\bibinfo{author}{\bibfnamefont{P.}~\bibnamefont{Morgan}},
  \bibinfo{journal}{Journal of Physics A: Mathematical and Theoretical}
  \textbf{\bibinfo{volume}{55}}, \bibinfo{pages}{254006}
  (\bibinfo{year}{2022}),
  \urlprefix\url{https://dx.doi.org/10.1088/1751-8121/ac6f2f}.

\end{thebibliography}

\end{document}